\documentclass[pdflatex,mathphys]{jnl}

\usepackage[normalem]{ulem}
\usepackage{wrapfig}
\usepackage{hyperref}
\usepackage[hyphenbreaks]{breakurl}

\jyear{2021}%

\theoremstyle{thmstyleone}%

\theoremstyle{thmstyletwo}%

\theoremstyle{thmstylethree}%

\begin{document}
\title[Nanoscale 3D tomography]{Nanoscale 3D tomography by in-flight fluorescence spectroscopy of atoms sputtered by a focused ion beam}

\author[1,2]{\fnm{Garrett} \sur{Budnik}}\email{garrett.budnik@student.uts.edu.au}

\author[1,5]{\fnm{John} \sur{Scott}}\email{john.scott@uts.edu.au}

\author[3]{\fnm{Chengge} \sur{Jiao}}\email{chengge.jiao@thermofisher.com}

\author[2]{\fnm{Mostafa} \sur{Maazouz}}\email{mostafa.maazouz@thermofisher.com}

\author[2]{\fnm{Galen} \sur{Gledhill}}\email{galen.gledhill@thermofisher.com}

\author[4,5]{\fnm{Lan} \sur{Fu}}\email{lan.fu@anu.edu.au}

\author[4,5]{\fnm{Hark Hoe} \sur{Tan}}\email{Hoe.Tan@anu.edu.au}

\author*[1,5]{\fnm{Milos} \sur{Toth}}\email{milos.toth@uts.edu.au}

\affil[1]{\orgdiv{School of Mathematical and Physical Sciences}, \orgname{University of Technology Sydney}, \orgaddress{\street{Ultimo}, \city{Sydney}, \postcode{2007}, \state{NSW}, \country{Australia}}}

\affil[2]{\orgdiv{Advanced Technology}, \orgname{Thermo Fisher Scientific}, \orgaddress{\street{NE Dawson Creek Dr.}, \city{Hillsboro}, \postcode{97124}, \state{OR}, \country{USA}}}

\affil[3]{\orgdiv{Applications Development}, \orgname{Thermo Fisher Scientific}, \orgaddress{\street{Achtseweg Noord 5}, \city{Eindhoven}, \postcode{5651 GG}, \country{Netherlands}}}

\affil[4]{\orgdiv{Department of Electronic Materials Engineering, Research School of Physics}, \orgname{The Australian National University}, \orgaddress{\street{Canberra}, \city{ACT}, \postcode{2601}, \country{Australia}}}

\affil[5]{\orgdiv{Australian Research Council Centre of Excellence for Transformative Meta-Optical Systems (TMOS)}}

\abstract{Nanoscale fabrication and characterisation techniques critically underpin a vast range of fields, including materials science, nanoelectronics and nanobiotechnology. Focused ion beam (FIB) techniques are particularly appealing due to their high spatial resolution and widespread use for processing of nanostructured materials and devices. Here, we introduce FIB-induced fluorescence spectroscopy (FIB-FS) as a nanoscale technique for spectroscopic detection of atoms sputtered by an ion beam. We use semiconductor heterostructures to demonstrate nanoscale lateral and depth resolution and show that it is limited by ion-induced intermixing of nanostructured materials. Sensitivity is demonstrated qualitatively by depth-profiling of 3.5, 5 and 8~nm quantum wells, and quantitatively by detection of trace-level impurities present at parts-per-million levels. To showcase the utility of the FIB-FS technique, we use it to characterise quantum wells and Li-ion batteries. Our work introduces FIB-FS as a high-resolution, high sensitivity, 3D analysis and tomography technique that combines the versatility of FIB nanofabrication techniques with the power of diffraction-unlimited fluorescence spectroscopy. It is applicable to all elements in the periodic table, and enables real-time analysis during direct-write nanofabrication by focused ion beams.}


\maketitle

\section{Introduction}\label{sec1}

Material characterisation techniques underpin progress in nano scale science and technology \cite{Lasagni, leng2013}. Elemental and chemical analysis techniques are particularly important, and their prevalence and usefulness are determined not only by basic performance metrics such as resolution, accuracy and sensitivity, but also by their applicability, in real time, to methods used to fabricate and process nanostructured materials and devices. The breadth of these requirements has led to regular use of numerous analysis methods as varied as atom probe tomography (APT), x-ray photoelectron spectroscopy (XPS), x-ray microscopy, scanning probe microscopy (SPM) and secondary-ion mass spectrometry (SIMS), as well as electron-beam techniques such as scanning transmission electron microscopy (STEM), electron energy loss spectroscopy (EELS), Auger electron spectroscopy (AES), cathodoluminescence (CL) spectroscopy, and energy-dispersive x-ray spectroscopy (EDX) \cite{Krivanek, Pattamattel, Sakdinawat, Bian, Gault, Zhisen, Polman, Egerton, Zhou, Grant, rinaldi2015}. However, in all cases, exceptional achievements in specific performance metrics are accompanied by drawbacks such as a highly restrictive sample geometry (e.g., in APT, STEM, and EELS), the need for an ultra-high vacuum environment (e.g., in AES), or the inability to detect hydrogen (e.g., in EDX, CL, AES and XPS). Consequently, there is continued interest in new approaches to nanocharacterisation, particularly ones that are inherently compatible with nanofabrication and processing methods \cite{Jun2017, Gamalski2012, Wirth2012}. 

\begin{figure*}[h!]
	\centering
	\includegraphics[width=\textwidth]{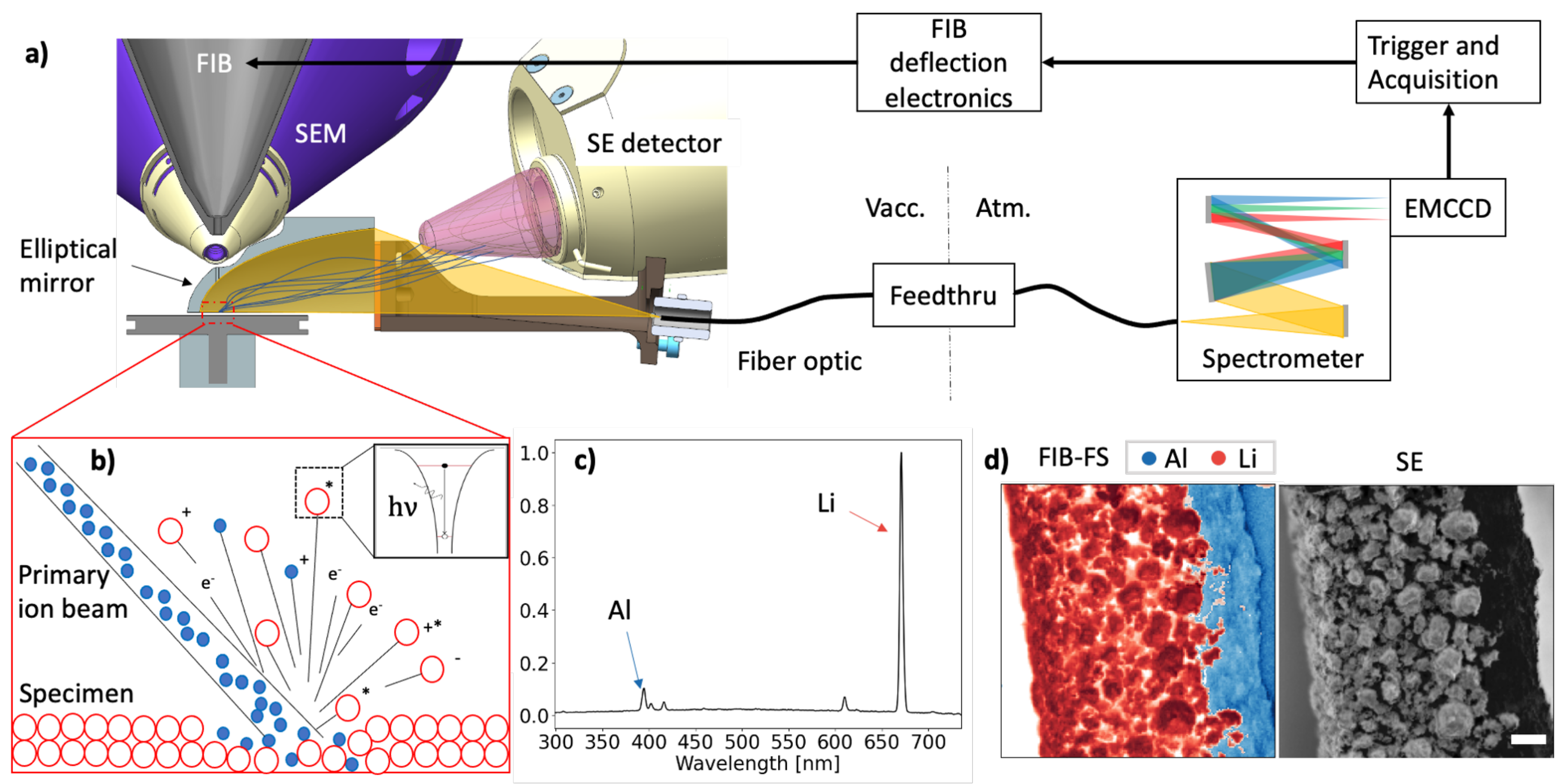}
	\caption{{\bf Nanoscale analysis by the FIB-FS technique. (a)} Schematic illustration of the FIB-FS setup implemented on an ion-electron dual beam microscope. Light emitted by sputtered atoms (orange) is collected by a retractable elliptical mirror (or a mini-lens, see SI Fig S1) located above the sample, coupled to a fiber and directed to an {\it ex-situ} optical setup. Electron imaging is performed using secondary electrons (SEs, blue trajectories) excited by the ion beam or the electron beam. {\bf (b)} Schematic illustration of FIB-induced particle emission. Ejected particles include excited sputtered neutrals and ions which relax through the emission of photons with energy $h\nu$. {\bf (c)} A FIB-FS spectrum showing Al and Li emissions corresponding to doublet transitions from the Al {$3s^24s (^2\text{S}_{\frac{1}{2}})$ excited state to the $3s^23p (^2\text{P}^{\text{o}}_{\frac{1}{2}, \frac{3}{2}})$ ground state (394.4~nm and 396.2~nm), and the Li $1s^22p (^2\text{P}^{\text{o}}_{\frac{3}{2},\frac{1}{2}})$ excited state to the $1s^22s (^2\text{S}_{\frac{1}{2}})$} ground state (670.77~nm and 670.8~nm). {\bf (d)} FIB-FS elemental map showing Al (blue) and Li (red) distributions in a Li battery cathode, and a corresponding SE image of the sample (scale bar = 10.6 $\mu$m).}
	\label{Fig1}
\end{figure*}

Here, we introduce a focused ion beam (FIB) analysis technique -- depicted in Figure \ref{Fig1}a-d -- based on in-flight fluorescence spectroscopy (FS) of atoms sputtered by energetic ions \cite{Thomas1979, Suchanska, Veje1989, Tsong1971, White1972, Tolk1977}. The FIB-FS technique enabled has high 3D spatial resolution, high sensitivity and the ability to detect all elements in the period table. Moreover, the technique is inherently compatible with FIB nanofabrication techniques, which have matured into versatile methods used broadly for direct-write processing of nanostructured materials and devices \cite{Narayan, Coskun, Haridy, Moll2018}. FIB-FS is complementary to SIMS which is, however, typically performed with highly specialised pulsed ion beam time-of-flight (TOF) analysis instruments \cite{Buchberger, Zhou, Bessette, Li2021, Collin, Loussert, Yang, Xu} that are inappropriate for nanofabrication applications. FIB-FS is also complementary to electron beam techniques such as EDX \cite{rinaldi2015} which are often employed on FIB nanofabrication systems like the one depicted in Figure \ref{Fig1}a, but have limitations that include an inability to detect hydrogen, and unfavorable spatial resolution and detection limits, as we demonstrate below.

\par We showcase FIB-FS as a robust, high-resolution, high sensitivity, 3D analysis and tomography technique. We study in detail the key performance metrics --- namely, lateral resolution, depth resolution and sensitivity --- and demonstrate the utility of FIB-FS using AlGaAs/GaAs quantum wells and Li-ion batteries. We achieve nano scale lateral and depth resolution, and show that it is limited fundamentally by ion-solid interactions. FIB-FS greatly enhances the nanoanalysis capabilities of focused ion beam systems and is inherently compatible with ion beam processing and nanofabrication techniques.

\section{Lateral Resolution}\label{sec2}

\begin{figure*}[h!]
	\centering
	\includegraphics[width=\textwidth]{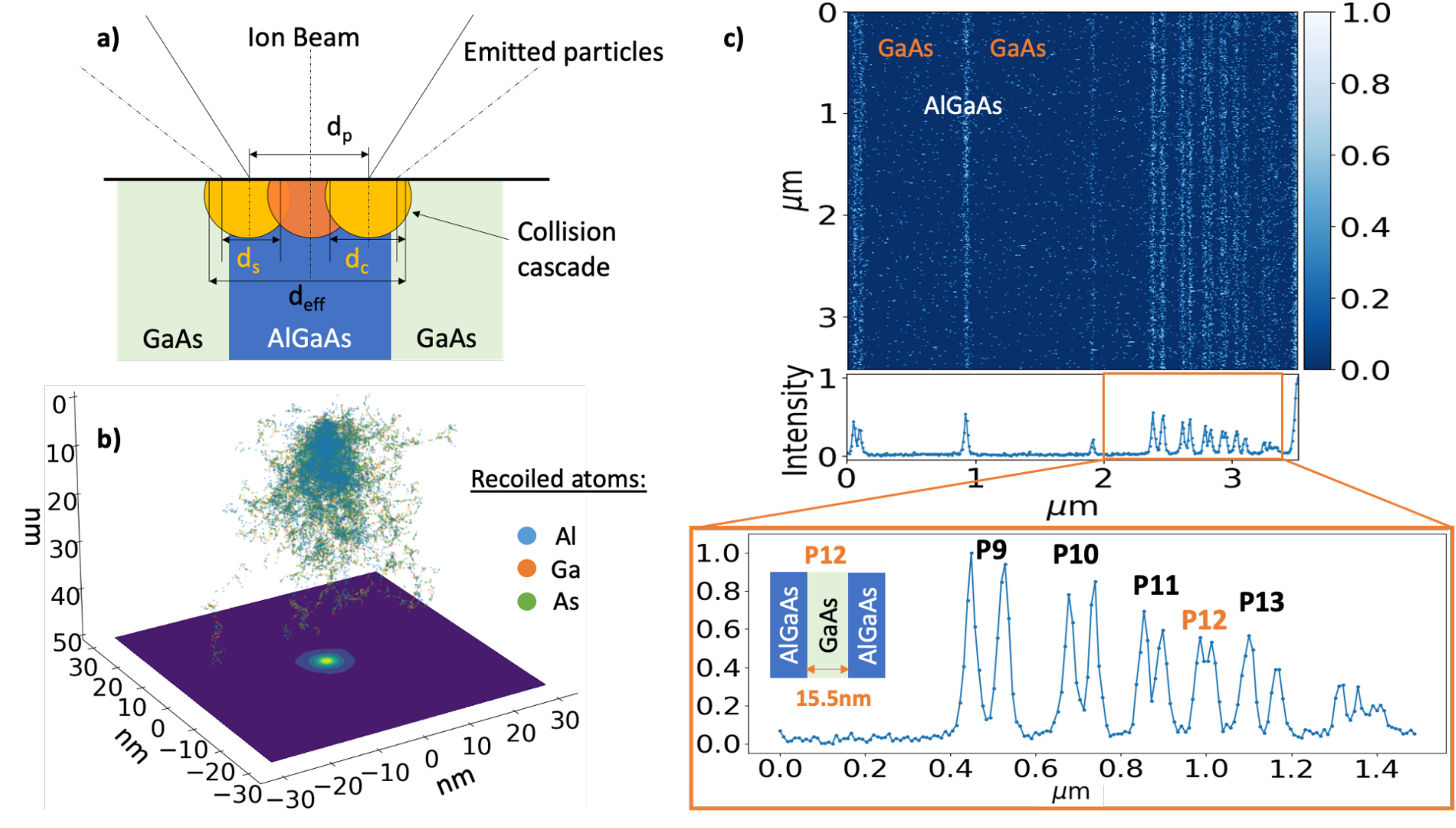}
	\caption{{\bf Lateral resolution of FIB-FS. (a)} Cross-sectional schematic of the effective probe diameter $\rm d_{eff}$, ion beam diameter $\rm d_p$, sputtered atom range $\rm d_s$, and collision cascade range $\rm d_c$. Each circle represents a collision cascade initiated by a single ion. The ion beam is incident onto a GaAs-AlGaAs-GaAs heterostructure. {\bf (b)} Monte Carlo trajectory simulation of 30~kV Ga+ ions incident onto AlGaAs. The plot shows the distributions of displaced Al, Ga and As atoms which are responsible for intermixing and broadening of interfaces in the sample. {\bf (c)} Aluminum elemental map of the BAM L200 resolution standard -- a heterostructure comprised of AlGaAs layers embedded in GaAs. The bottom panel is the corresponding integrated line spread function, showing a resolution of 15.5~nm through AlGaAs-GaAs-AlGaAs stripe pattern P12, depicted in the inset.}
	\label{Fig2}
\end{figure*}

We start with an analysis of spatial resolution. The FIB-FS technique is based on fluorescence spectroscopy of sputtered particles. Lateral resolution is therefore expected to be limited technologically by the ion beam diameter and fundamentally by ion-induced collision processes in the sample, as is the case in other ion beam techniques \cite{Betz1991, Pillatsch}. Figure \ref{Fig2}a illustrates schematically the three factors that determine the FIB-FS probe diameter $\text{d}_e$: the ion beam diameter $(\text{d}_p)$, the diameter of the region from which atoms are sputtered as a result of ion impact $(\text{d}_s)$, and the diameter of the volume in which atoms are displaced and intermixed in the sample $(\text{d}_c)$. The latter is a consequence of collision cascades initiated by each ion and causes intermixing and broadening of interfaces in samples such as the GaAs-AlGaAs-GaAs heterostructure depicted in Figure \ref{Fig2}a. Collision cascades determine the size of the ion-solid interaction volume, which is shown in Figure \ref{Fig2}b for the case of an ideal (zero diameter) 30~keV Ga+ beam incident onto AlGaAs.
\par In order to measure lateral resolution, we use a certified standard sample (BAM L200) which is commonly used to characterize FIB-SIMS performance \cite{Senoner2004, Senoner2015, Senoner2007, Pillatsch, Whitby}. The sample is a cross-sectioned, epitaxially-grown AlGaAs-GaAs heterostructure with a range of layer thicknesses and periods down to a few nm (see SI, Figure S2 and Table S1). Figure \ref{Fig2}c shows a FIB-FS image of the sample generated by summing the 394.4 and 396.2~nm Al emissions shown in SI Figure S3. Bright stripes in the map correspond to AlGaAs layers embedded in GaAs. The bottom panel is an integrated line profile of the Al map (i.e., the line spread function), showing AlGaAs-GaAs-AlGaAs stripe patterns P9--P13, in which the GaAs layer thickness varies from 38.3~nm to 11.5~nm. The thinnest GaAs layer resolved in the line profile is in pattern P12. It has a thickness of 15.5~nm, which corresponds to the resolution of the FIB-FS map.
\par The employed 30~keV ion beam has a diameter $\text{d}_p$ of $\sim$10~nm (see SI Section S3, Figures S5 and S6), which is approximately 6~nm smaller than the measured resolution. To confirm that the difference is due to ion-induced collision processes, we used the Monte Carlo code SRIM \cite{Ziegler1977} to simulate the trajectories of the incident Ga+ ions and recoiled atoms in AlGaAs (Figure \ref{Fig2}b). To simulate sputtering and calculate $\text{d}_s$, recoiled atoms are tracked in the sample and their final coordinates are recorded at the surface before ejection. In the case of 30~keV Ga+ ions incident onto AlGaAs, most sputtered atoms are ejected from within a 10~nm radius (see SI Fig S4). The standard deviation of the lateral distribution (i.e., the sputtered atom range $\text{d}_s$) is $\rm \sim 4.3~nm$. Similarly, to determine $\text{d}_c$, the Ga ions are tracked and their coordinates are recorded when they come to rest in the sample. The mean lateral range of the implanted ions is given by $\text{r}_l = \sum_i \text{y}_i/N$, where $\text{y}$ is the projection of the coordinates in the plane of the sample and $N$ is the number of ions simulated \cite{Zieglersetup}. For 30~keV Ga+ ions in AlGaAs, $\text{r}_l \sim $5.6~nm, which corresponds to the collision cascade range $\text{d}_c$. Hence, the sputtered atom range limits lateral resolution to $\rm\sim 4~nm$ and intermixing to $\rm\sim 6~nm$. Given the ion beam diameter of $\rm\sim 10~nm$, this is in excellent agreement with the measured resolution of $\rm\sim 16~nm$ seen in Figure \ref{Fig2}c. Moreover, this analysis shows that, fundamentally, lateral resolution of this fluorescence spectroscopy technique is limited by ion-solid interactions to length scales well beyond the optical diffraction limit defined by the wavelength of the detected light. This makes FIB-FS highly appealing as a nanoscale analysis technique that is intrinsically compatible with FIB-based nanofabrication methods.
\par We note that the ion beam diameter is limited by the virtual size of the ion source and the ion focusing optics \cite{orloff2009}. In general, $\rm d_p$ decreases with ion beam energy and reciprocal current, and it is a function of the ion species. These limitations are technological and improve continuously in modern FIB instruments \cite{Bassim2014}. Conversely, the limitations imposed by sputtering and intermixing are fundamental. They are determined by the ion mass and energy, and by the masses and binding energies of atoms that make up the sample. However, systematic studies of the effects of ion mass and energy on lateral resolution are complicated by the fact that they affect not only $\rm d_s$ and $\rm d_c$, but also $\rm d_p$. This is, however, not the case for depth resolution, which is independent of $\rm d_p$. Hence, we next study depth profiling in order to elucidate the fundamental effects of ion species and energy on FIB-FS spatial resolution.

\section{Depth Resolution and 3D Tomography}\label{sec3}

\begin{figure*}[b]
	\centering
	\includegraphics[width=\textwidth]{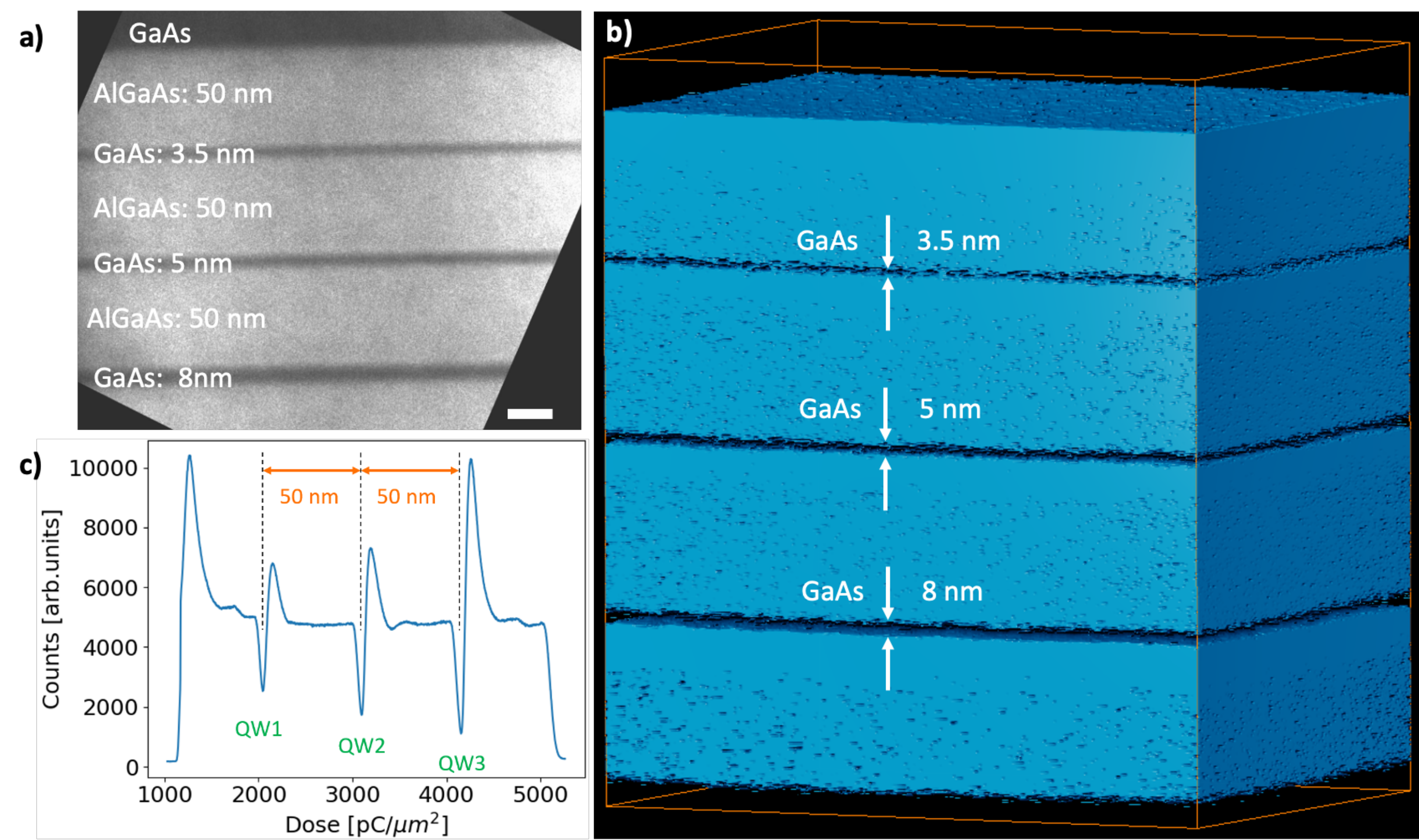}
	\caption{{\bf FIB-FS depth-profiling and 3D imaging. (a)} TEM image of a heterostructure comprised of three GaAs quantum wells in AlGaAs (scale bar = 20 nm). The quantum well thicknesses are 3.5, 5 and 8~nm. {\bf (b)} 3D volume reconstruction of the heterostructure generated by summing, at each pixel, the FIB-FS Al emissions at 394.4~nm and 396.2~nm (2~keV Ga$^+$ beam, 3.2~nA, 1.68~pC/$\mu$m$^2$ per slice). All three GaAs quantum wells are clearly resolved in the image (the lateral field width is $\rm 50~\mu m$).{\bf (c)} Corresponding FIB-FS depth line profile plotted as a function of ion dose per unit area. A minimum in the Al signal level at each of the three GaAs quantum wells is marked by a dashed line. Two 50~nm scale bars are defined by the dose intervals between the minima.}
	\label{Fig3}
\end{figure*}

\par To demonstrate FIB-FS depth profiling, we employ a sample comprised of three GaAs quantum wells embedded in AlGaAs barriers grown epitaxially by metalorganic chemical vapour deposition \cite{Tan1996}. A TEM image of the sample is shown in Figure \ref{Fig3}a. The quantum wells have thicknesses of 3.5, 5 and 8~nm. Figure \ref{Fig3}b is a 3D FIB-FS image of the sample generated by summing, at each pixel, the FIB-FS Al emissions at 394.4~nm and 396.2~nm (see SI Fig S3). The 3D image stack consists of 2,467 FIB-FS maps, each collected in a single scan lasting 1.13~s. The lateral field width is 50 $\mu$m. The first 600 frames were removed (to ignore most of a GaAs capping layer), and each frame was cropped by 1 $\mu$m at the outer rim to remove crater edge/wall effects \cite{Magee1982}. All three quantum wells are clearly resolved in the 3D FIB-FS image, illustrating qualitatively the outstanding sensitivity and resolving power of the technique. 
\par We expect FIB-FS depth resolution to be limited by intermixing of atoms caused by ion impacts. This is illustrated by the Al depth profile shown in Figure \ref{Fig3}c, obtained by integrating the image stack in Figure \ref{Fig3}b. Each GaAs quantum well is again well resolved in the depth profile. It is seen as a dip followed by a peak in the Al signal level -- a result of knock-on displacement of atoms along the beam axis and preferential sputtering of Al relative to Ga and As. The dip and peak magnitudes are determined by ion mass and energy, layer thicknesses and sputter rates. This is typical of collision kinetics at interfaces and it is documented thoroughly in SIMS literature \cite{Whittmaack1982, Whittmaack1990, Whittmaack1997}. Additional analysis of FIB-FS depth profiling is provided in the SI (Section S4), where we elucidate further the broadening of GaAs-AlGaAs interfaces and highlight a distinction between qualitative sensitivity (illustrated here by the ability to detect a 3.5~nm quantum well) and quantitative depth resolution.

\par In order to quantify depth resolution and measure dependencies on ion mass and energy, we employed NIST Standard Reference Material 2135c, a sputter depth calibration standard comprised of alternating Cr and Ni layers (see SI Figure S7 and S8). Details of the quantification procedure can be found in SI Section S4 and Figure S9. Briefly, a 1D depth profile is measured across a Ni-Cr interface by collecting the Cr FIB-FS signal as a function of ion dose, and resolution is defined as the width of the measured interface. Specifically, resolution is the full-width-at-half-maximum (FWHM) of the first derivative of the line spread function corresponding to a Cr FIB-FS signal rise time of 12\%--88\%.

Starting with 2~keV Ga+ ions, the interface width measured by FIB-FS is 12.8~nm. This is in excellent agreement with a reference measurement of 12.2~nm obtained from an AES depth profile produced by sputtering the sample with 1~keV Ar+ ions (the value of $\sim$12~nm corresponds to the true width of the interface between the Ni and Cr layers, as is discussed in the SI, Section S4). Increasing the Ga+ ion energy to 30~keV reduced the measured depth resolution to $\sim$35~nm, and changing the ion species to 30~keV Xe+ increased it to $\sim$26~nm. These trends are expected since the ion range increases with energy and decreases with ion mass, unless the implanted ions are chemically active and modify the binding energies of atoms in the sample. Such a chemical effect is demonstrated in the SI, Section S4, Figure S10, by an O+ beam which causes oxidation of the sample. Oxidation increased the binding energies of atoms in the sample, and thereby reduced the ion penetration range and increased the measured FIB-FS depth resolution to $\sim$23~nm at 30~keV.

\section{Sensitivity and detection limits}\label{secS}
\par We now turn to sensitivity and detection limits of the FIB-FS technique. All elements in the periodic table are fluorescent, and detection of trace elements present in the parts-per-million range should be feasible based on prior measurements of fluorescence yields of sputtered atoms \cite{Tsong1983, White1972}. To confirm this using a highly focused ion beam, we adapted methods established in the SIMS literature \cite{Chakraborty, Ottolini}. Specifically, we used calibration standards of known compositions to convert photon counts to atomic concentrations. Details are provided in the SI Section S5 where we measure FIB-FS potassium and lithium minimum detection limits of 3.9 ppm and 0.8 ppm, respectively. These results demonstrate that FIB-FS is highly competitive as a method for the detection of trace-level impurities. 

We note that the above values are not universal. Photon emission efficiencies are influenced by sample matrix effects and electron transfer processes that vary with surface electronic structure, vacuum pressure, incident ion energy, angle, and flux \cite{Thomas1979, Veje1988, Berthold1997, Cortona1999}. In nanostructured materials, sensitivity also depends on ion dose, and is limited ultimately by sputtering which removes the material that is being analysed. Moreover, for matrices in which the atomic density is known, the sensitivity ($n_i$) of element $i$ can be expressed as a function of the sputtered volume $V$ \cite{Pillatsch}, $n_i = \frac{P_i}{V \beta_i}$, where $P_i$ is photon counts and $\beta_i$ is the photon yield per sputtered atom. That is, sensitivity and sputtered volume are inter-related. In the case of nanostructured materials, sensitivity is therefore a function of feature size and the required spatial resolution. In general, ultimate FIB-FS sensitivity is determined by the light collection efficiency, the ion beam parameters (beam current, exposure time, etc.), the concentrations of elements with nearby spectral peaks (dynamic range), the required spatial resolution, and the photon yield per sputtered atom. The highest photon yields occur in light alkali and alkaline earth metals \cite{Tsong1983}, making the technique appealing for the applications like Li ion battery research and complementary to techniques such as EDX, a technique implemented on most ion-electron dual beam microscopes.

\section{Lithium Ion Battery Analysis} \label{sec4}

\begin{figure*}[h!]
	\centering
	\includegraphics[width=\textwidth]{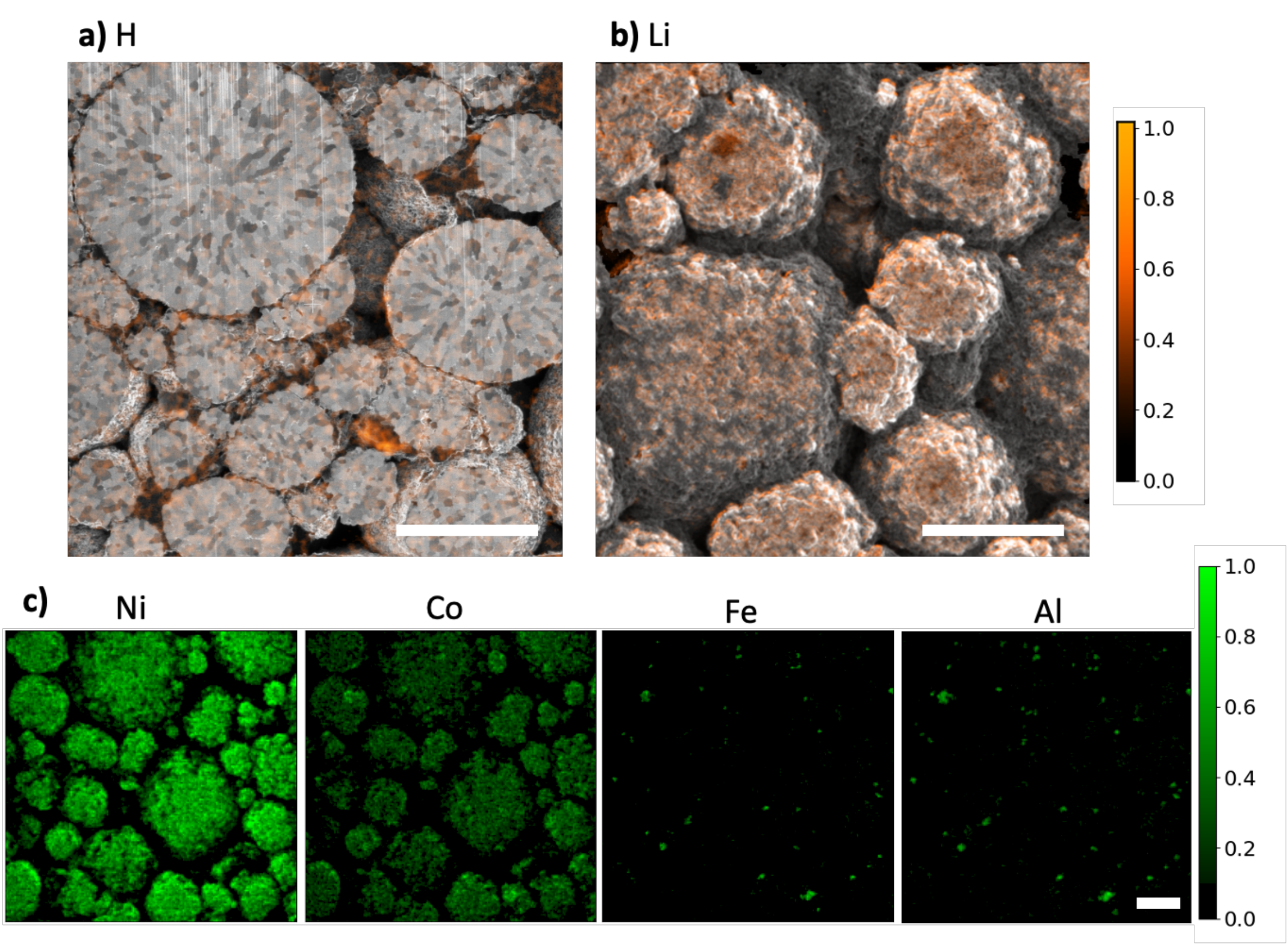}
	\caption{{\bf FIB-FS elemental maps of an NMC cathode. (a)} Hydrogen map (orange) overlaid on a secondary electron image (grey) of a cross-section that was prepared using the Ga$^+$ focused ion beam (scale bar = 10 $\mu$m, 30 keV Ga$^+$, 1.07 nC/$\mu$m$^2$, H emission at 656.3~nm). {\bf (b)} Lithium map (orange) overlaid on a top-down secondary electron image (grey) of the sample: (scale bar = 6 $\mu$m, 30 keV Ga$^+$, 51.9 pC/$\mu$m$^2$, Li emissions at 670.77 nm and 670.8 nm, see SI Section S7, Figure S17). {\bf (c)} FIB-FS maps acquired in parallel from a single region of the cathode showing distributions of the mixed metal base of Ni and Co as well as the trace metal dopants Fe and Al (scale bar = 6 $\mu$m, 30 keV Ga$^+$ 518 pC/$\mu$m$^2$). Color scales are normalized photon counts (arb. units).}
	\label{Fig4}
\end{figure*}

Finally, we use a lithium ion battery cathode to demonstrate the utility of the FIB-FS technique and show how it is complimentary to EDX. The sample is a lithium-nickel-manganese-cobalt-oxide (NMC) cathode of the form Li(Ni$_x$Mn$_y$Co$_z$M’$_m$)O$_2$, where M’ denotes dopants of Fe, Al, Zr, and Mg at 0.02 at.\% and $x+y+z+m = 1$. EDX yields a good understanding of the general chemical composition, as is shown in the SI Section S6, Figure S15. However, EDX analysis of light elements is challenging due to peak overlaps at low x-ray energies, electron-driven buildup of surface contaminants that contain carbon and oxygen, and a need for windowless detectors for efficient transmission of low energy x-rays \cite{Hovington, Burgess}. Moreover, detection of hydrogen, a ubiquitous dopant of interest in many materials is not possible by EDX. Conversely, FIB-FS has no inherent limitations for spectroscopy of light elements, as is illustrated by the hydrogen map of a cross-section of the NMC cathode shown in Figure \ref{Fig4}a. The map is overlaid on a secondary electron image which reveals crystallinity of the cross-sectioned metallic grains. Similarly, a lithium FIB-FS map is shown in Figure \ref{Fig4}b. This sample region was not cross-sectioned, and the corresponding electron image reveals the surface topography of the analysed sample region. 

\par NMC cathodes contain trace amounts of metals that affect performance. For example, aluminum is used to increase initial capacity,  iron denotes anti-site defects and magnesium is generally associated with enhanced structural stability \cite{Hashem, Lipson}. Spatial distributions of these dopants are therefore of interest and depicted by the FIB-FS maps in Figure \ref{Fig4}c. Ni and Co maps show the mixed base metal grains, whilst the Fe and Al maps show the distributions of these dopants within these grains (an Mg map is shown in SI Figure S18 together with a depth profile that shows the 3D dopant distribution within the cathode). Mn maps are ignored due to the presence of background Ga emission (see SI Section S8).

The maps in Figure \ref{Fig4}c illustrate some sample-specific benefits of FIB-FS that make it complementary to EDX. These include an improved signal-to-noise ratio (SNR) and spatial resolution (reference EDX maps are shown in SI Figure S15). The difference in resolution is inherent to electron-solid and ion-solid interactions that generate the EDX and FIB-FS signals, respectively. Due to their low mass, the electron range is on the order of 1~$\rm\mu m$ at 10~keV, whereas that of 10~keV Ga$^+$ ions is on the order of 10~nm, resulting in higher resolution as well as greater surface sensitivity.

\section{Conclusion} \label{sec6}
FIB-FS combines the favorable properties of focused ion beam nanofabrication techniques with the power of diffraction-unlimited spectroscopic analysis. Through the design and deployment of an ultra-sensitive optical detection scheme, we have demonstrated the fundamental characteristics of FIB-FS by 3D tomography, quantitative analyses of nanoscale lateral and depth resolution, detection of trace-level impurities and demonstrations of applicability to GaAs/AlGaAs quantum wells and lithium ion batteries. Nonetheless, the FIB-FS technique is in its infancy. As it matures, it is reasonable to expect improvements in performance, and in understanding of the electronic excitation mechanisms that lead to fluorescence of atoms sputtered by energetic ions. To this end, FIB-FS is both a nanoscale analysis technique and a means for exploring fundamental interactions between ions and matter.

\section{Methods}\label{sec7}
All experiments were performed using two Thermo Fisher Scientific coincident FIB-SEM systems. Ga$^+$ beam experiments were done using a Helios 5 LMIS system while the Xe$^+$ and O$^+$ experiments in the SI were performed using a Hydra plasma FIB system.

\par To collect photons emitted by sputtered atoms, we implement a spectroscopic technique incorporating an ellipsoidal mirror designed with semimajor and semiminor axes to couple light from its focal point at the incident region under the FIB/SEM to a fiber optic located at the opposing focal point. One side of the mirror is left open to allow secondary electron and secondary ion escape for imaging when placed over the sample. A slotted aperture is drilled on the column end from 0 to 53 degrees from the zenith to allow for both FIB and SEM imaging at normal incidence. The mirror is cut to achieve a 73\% solid angle collection efficiency when placed 600 $\mu$m above the sample surface and is positioned with a non-magnetic 3-axis substage system mounted on the main chamber stage system. Nanometer precision of the substages allow for fine adjustment of the mirror, while the main stage mounting allows for full tilt between FIB and SEM normal incidence angles (0-52$^\circ$).
\par A large core multimode fiber directs the coupled light out of the vacuum chamber through a custom fiber feedthrough flange to either an Kymera 193i spectrograph equipped with a back-illuminated electron multiplying CCD (EMCCD) iXon 888 for spectral decomposition or to a Hamamatsu R9880U PMT for imaging. Different modes of collection are used to enable various modes of operation:
\begin{enumerate}
    \item Survey Mode: the FIB scan and EMCCD camera acquisition time are decoupled to allow an average spectrum of a region of interest scanned with the ion beam.
    \item Hyperspectral Mode: the FIB scan and EMCCD camera acquisition time are clocked in sequence such that the dwell time of the FIB pattern equals the exposure time of the camera, resulting in a full spectrum collected at every pixel. The spectra are then used to generate elemental maps.
    \item Photon Imaging Mode: the FIB scan and the PMT acquisition are clocked the same as (2), but the light is directed towards a PMT. The PMT allows for ns and $\mu$s dwell times, resulting in real-time high resolution photon imaging. The light path optionally employs filters to isolate specific emissions and enable high throughput elemental mapping. 
\end{enumerate}
An example is provided in the SI, section S7. The 3D reconstructions presented in this work consist of stacks of multiple 2D maps collected in series using either Hyperspectral Mode or Imaging Mode.

Care is taken to prepare the BAM L200 lateral resolution sample. A thin native oxide was present on the surface of the sample that impacts secondary electron imaging of the vertical strips, so a gentle clean using a 5~keV defocused Ga beam (2nA, 50 ns dwell, 90\% overlap, 100 $\mu$m defocus, 5 second irradiation) was used to remove it before imaging. A similar process was used for depth profiling of the Ni/Cr standard.

\backmatter

\bmhead{Supplementary information}

This article contains supplementary information.

\bmhead{Acknowledgments}

The authors acknowledge financial support from the Australian Research Council (CE200100010, LP170100150, DP190101058). The authors thank Tom Nichols and Keith Ragsdale for contributions to mechanical engineering of hardware used in this work. The Australian National Fabrication Facility is acknowledged for access to nanofabrication facilities used in this work.

\section*{Ethics declarations}
The authors declare no competing interests.

\bibliography{Main_bibfile}

\end{document}


\title[Supporting Information]{Supporting Information for: Nanoscale 3D tomography by in-flight fluorescence spectroscopy of atoms sputtered by a focused ion beam}

\author[1,2]{\fnm{Garrett} \sur{Budnik}}\email{garrett.budnik@student.uts.edu.au}

\author[1,5]{\fnm{John} \sur{Scott}}\email{john.scott@uts.edu.au}

\author[3]{\fnm{Chengge} \sur{Jiao}}\email{chengge.jiao@thermofisher.com}

\author[2]{\fnm{Mostafa} \sur{Maazouz}}\email{mostafa.maazouz@thermofisher.com}

\author[2]{\fnm{Galen} \sur{Gledhill}}\email{galen.gledhill@thermofisher.com}

\author[4,5]{\fnm{Lan} \sur{Fu}}\email{lan.fu@anu.edu.au}

\author[4,5]{\fnm{Hark Hoe} \sur{Tan}}\email{Hoe.Tan@anu.edu.au}

\author*[1,5]{\fnm{Milos} \sur{Toth}}\email{milos.toth@uts.edu.au}

\affil[1]{\orgdiv{School of Mathematical and Physical Sciences}, \orgname{University of Technology Sydney}, \orgaddress{\street{Ultimo}, \city{Sydney}, \postcode{2007}, \state{NSW}, \country{Australia}}}

\affil[2]{\orgdiv{Advanced Technology}, \orgname{Thermo Fisher Scientific}, \orgaddress{\street{NE Dawson Creek Dr.}, \city{Hillsboro}, \postcode{97124}, \state{OR}, \country{USA}}}

\affil[3]{\orgdiv{Applications Development}, \orgname{Thermo Fisher Scientific}, \orgaddress{\street{Achtseweg Noord 5}, \city{Eindhoven}, \postcode{5651 GG}, \country{Netherlands}}}

\affil[4]{\orgdiv{Department of Electronic Materials Engineering, Research School of Physics}, \orgname{The Australian National University}, \orgaddress{\street{Canberra}, \city{ACT}, \postcode{2601}, \country{Australia}}}

\affil[5]{\orgdiv{Australian Research Council Centre of Excellence for Transformative Meta-Optical Systems (TMOS)}}

\maketitle

\section{Light Collection Hardware}
\begin{figure*}[b!]
	\centering
	\includegraphics[width=\textwidth]{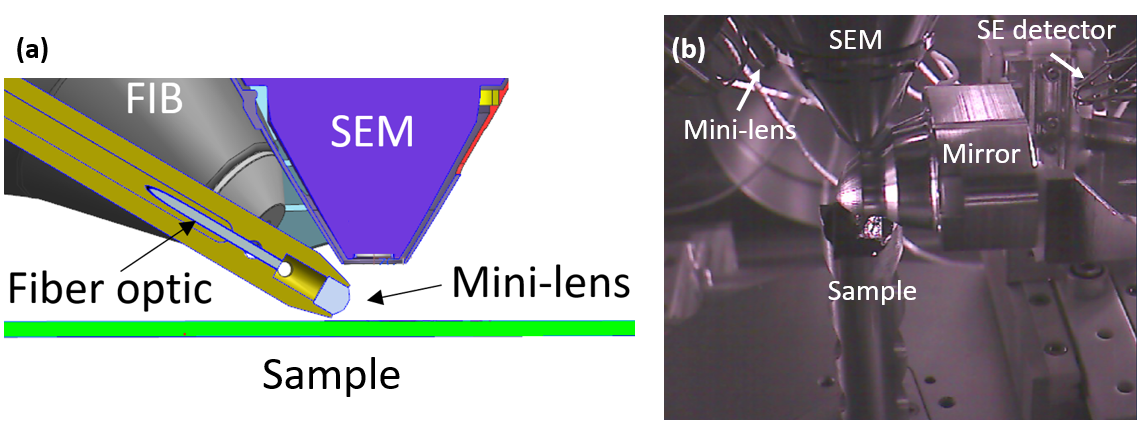}
	\caption{In-chamber light collection setups used for FIB-FS. {\bf (a)} Cross-sectional schematic of a mini-lens optic that incorporates a retractable  rod bored with a 1.8~mm hole which allows for a multimode fiber optic cable to be fed through to a lens system on the distal end. The solid angle is $\sim$10\% and the fiber coupling efficiency is $\sim$70\%. {\bf (b)} In-chamber photograph showing an elliptical mirror located below the pole pieces of the ion beam and electron beam optical columns. A cross-sectional schematic of the setup is shown in Figure 1a of the paper.}
	\label{SI1}
\end{figure*}

Schematics of two light collection setups that we used for focused ion beam-induced fluorescence spectroscopy (FIB-FS) are shown in Figure \ref{SI1}. Each setup couples the collected light to a multimode fiber, through the specimen vacuum chamber wall, and into the optical paths described in the Methods sections.

\par The two setups are optimised for compactness and light collection efficiency, respectively. The mini-lens setup in Figure \ref{SI1}a allows greatest flexibility in use of other hardware such as electron and ion detectors and gas injectors positioned around the sample. The setup depicted in Figure \ref{SI1}a shows a 2~mm diameter half-drums lens. We also employed ball lenses, half-ball lenses, and various combinations of each (depending on the fiber optic). Best performance was achieved using a single half-drum lens coupled to a 1~mm core multimode fiber, yielding the best overall efficiency (solid angle and coupling) of $\sim 4.5\%$.

The ultra-high efficiency elliptical mirror in Figure \ref{SI1}b occupies a larger volume above the sample, but has a large solid angle of collection of 73\%. Both the mini-lens and the mirror shown in Figure \ref{SI1} are retractable and can be interchanged during FIB-FS workflows. 

\section{Lateral Resolution}

\begin{figure*}[h!]
	\centering
	\includegraphics[width=\textwidth]{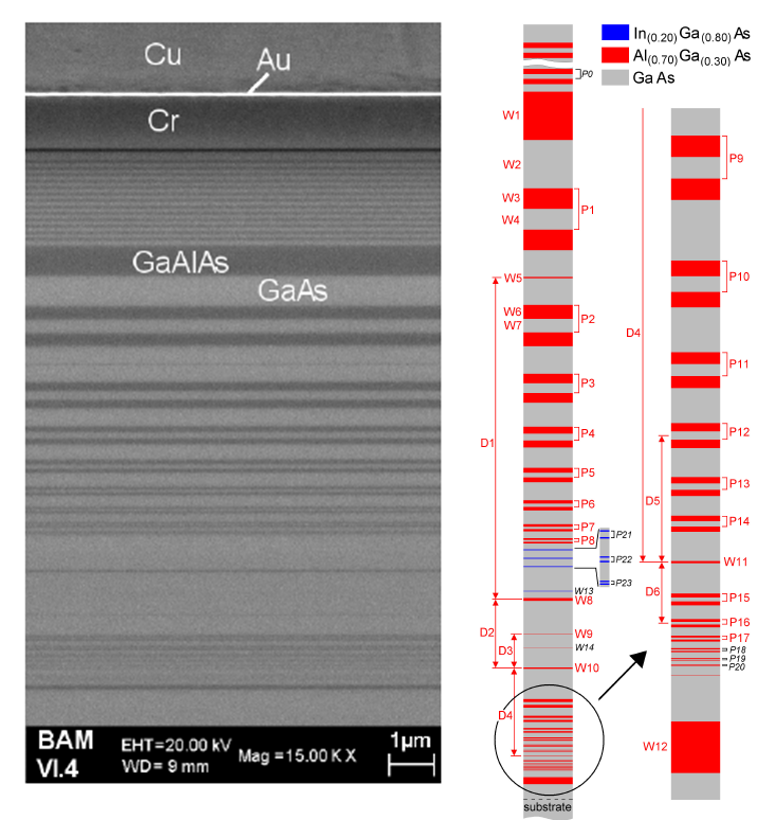}
	\caption{SEM image and schematic of the BAM L200 resolution calibration sample. It consists of nanoscale stripe patterns with gratings of alternating GaAs and AlGaAs with varying periods for lateral resolution testing and calibration. The layer stack was grown by Metal Organic Vapor Phase Epitaxy. Layer thicknesses were certified by high resolution X-ray diffraction, optical reflection spectroscopy, and transmission electron microscopy. Image provided in the certification data sheet \cite{Reiners}.}
	\label{SI2}
\end{figure*}
\begin{table}[h!]
	\centering
	\includegraphics[width=\textwidth]{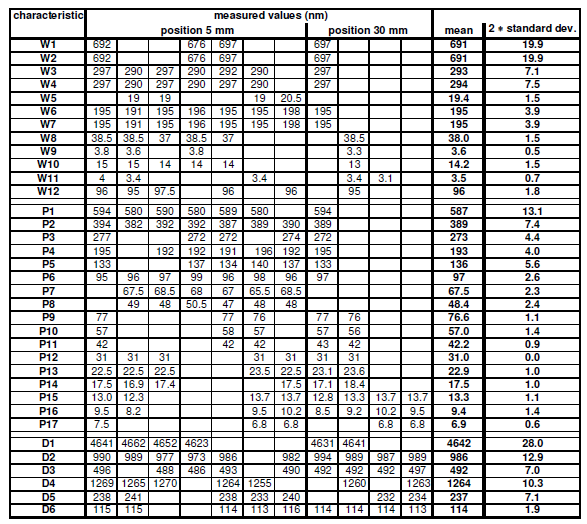}
	\caption{Table of certified values in nanometers for the features on BAM L200 sample reported in the certificate and their corresponding locations on the sample. In our experiments, we imaged a region of the sample containing strip periods P7 through single line W12. Table provided in the BAM-L200 certification report \cite{Senonercert}.}
  \label{SIT1}
\end{table}

\par BAM L200 is a certified resolution standard from Bundesanstalt f\"{u}r Materialforschung und -pr\"{u}fung (BAM) used broadly for sputter-based measurements \cite{Senoner2007, Senoner2015, Senoner2004, Pillatsch}. It consists of strip widths W, grating periods P, and strip distances D designed for various resolution measurement methods: knife-edge, image modulation of a square wave grating, full width at half maximum (FWHM) of a single strip, or strip-to-strip distance. An SEM image alongside the illustrative pattern map is shown in Figure \ref{SI2} and depicts the orientation of the sample used to measure spatial resolution in this study. Metrics corresponding to the various patterns and periods are listed in Table \ref{SIT1}.

\begin{figure*}[h!]
	\centering
	\includegraphics[width=\textwidth]{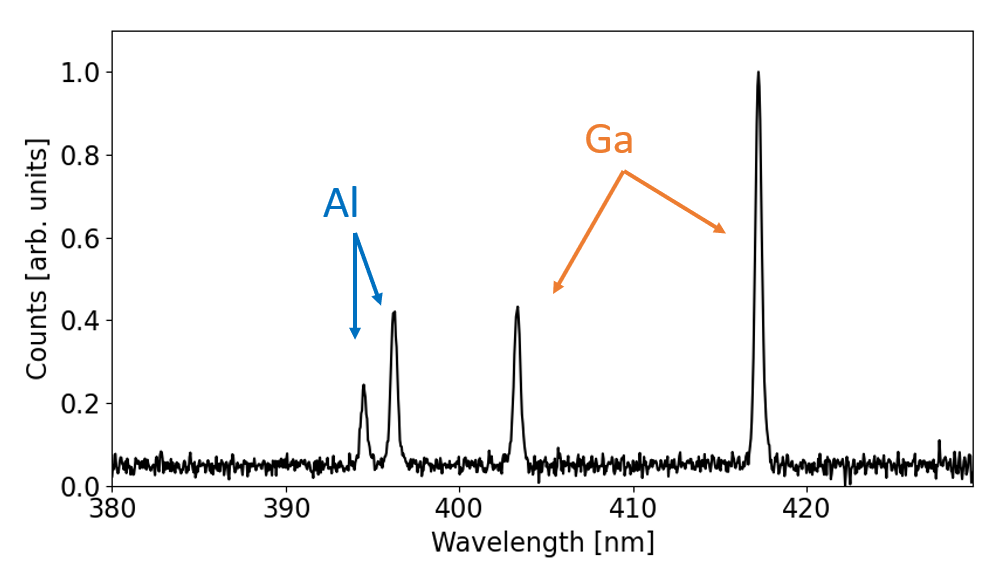}
	\caption{FIB-FS spectrum of AlGaAs showing the Al doublet transition from the $4s$ excited state to the $3p$ ground state at 394.4 nm and 396.2 nm and the Ga doublet transition from the $5s$ excited state to the $4p$ ground state at 403.3 nm and 417.2 nm. Due to the presence of Ga in each heterostructure and in the primary ion beam, only the Al signal was used for mapping in both the lateral resolution and depth profiling measurements.}
	\label{SI3}
\end{figure*}
\par FIB-FS resolution measurements were performed by detection of the Al doublet corresponding to transitions from the $3s^24s\, (^2\text{S}_{\frac{1}{2}})$ excited state to the $3s^23p \,(^2\text{P}^\text{o}_{\frac{1}{2},\frac{3}{2}})$ ground state, located at 394.4~nm and 396.2~nm, respectively. The emission lines are labelled in Figure \ref{SI3} –- a typical FIB-FS spectrum of an AlGaAs sample. The Al map in the paper is an average of 256 images, each comprised of 512x442 pixels, obtained using a 30 keV, 7 pA Ga+ FIB, and a dwell time of 500 ns, corresponding to a total dose of 11.6 pC/\textmu m$^2$.

\begin{figure}
	\centering
	\includegraphics[width=\textwidth]{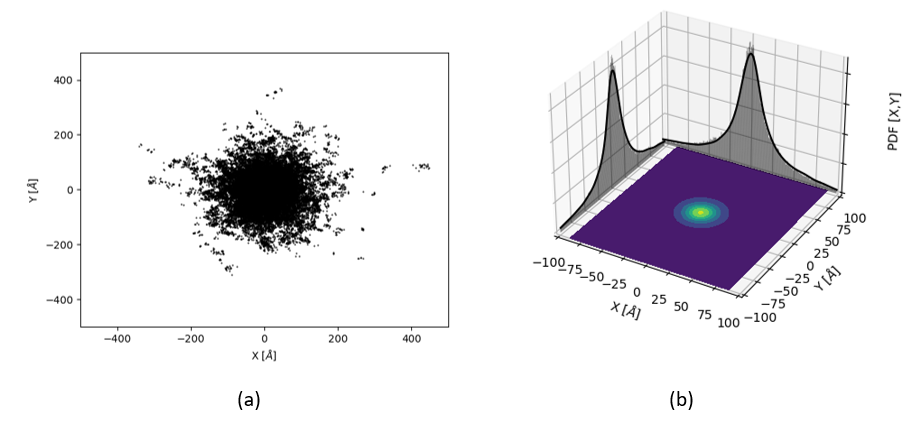}
	\caption{SRIM Monte Carlo simulation of 10,000 Ga+ ions (30 keV) incident on AlGaAs. For sputtered atoms, the program records their final coordinates before ejection. {\bf (a)} Top-down scatter plot of the range of sputtered particle ejection locations. {\bf (b)} Gaussian kernel density estimation (KDE) with a bin of roughly 1 angstrom was used to derive the probability density function which is projected alongside the contoured KDE surface. Here, the vast majority of sputtered particles originate from a radius of $\sim 10$~nm.}
  \label{SI5}
\end{figure}

\par Both the sputtered atom range $\rm (d_s)$, and collision cascade range $\rm (d_c)$ were investigated to determine their contributions to lateral resolution limits. As is discussed in the paper, the Monte Carlo program tracks the locations of sputtered atoms as well as locations of lattice collisions and recoils. Figure \ref{SI5} shows the results of a simulation of 30 keV Ga+ incident on AlGaAs used to calculate $\rm (d_s)$, which was found to be smaller than $\rm (d_c)$. The latter is therefore the limiting factor in spatial resolution.

\section{Ga FIB Spot Size}
\begin{figure}[h!]
	\centering
	\includegraphics[width=\textwidth]{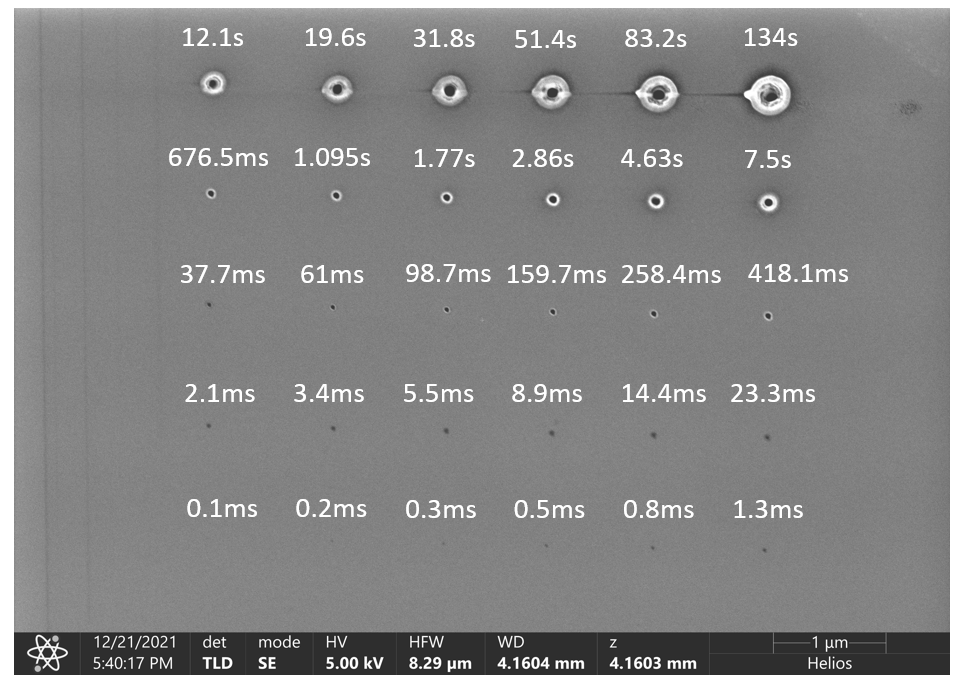}
	\caption{SEM secondary electron image of a spot exposure grid generated in Si by a 7.5 pA, 30 keV Ga$^+$ ion beam. The spot exposure time increases as a Fibonacci sequence from the bottom left to the top right, starting at 0.1~ms. Note: the patterning framework of this FIB instrument has a maximum exposure time of 100 ms so streaks are seen to the left of the high dwell spots due to ion beam blanking.}
  \label{SIspotburn}
\end{figure}

Our analysis of lateral resolution requires knowledge of the FIB diameter $\rm (d_p)$ which can, in principle, be measured using images of spots generated on a sample by a stationary ion beam. However, in the case of highly focused beams, such measurements are a strong function of the FIB exposure time, as is illustrated by Figure \ref{SIspotburn} --- an SEM secondary electron image of a grid of spots generated by the Ga+ ion beam used for our resolution measurements. In the limit of short exposure time, ($\sim$0.1~ms) the ion beam does not generate sufficient image contrast to be detected in secondary electron images. Conversely, relatively long dwell times result in significant surface topography which can be measured, but the pit shapes are not representative of the beam shape due to re-deposition of sputtered atoms at pit sidewalls --- a problem that is negligible in the case of broad ion beams, but significant in the case of the highly focused beams used in our experiments. Consequently, we measured $\rm d_p$ by analysis of contrast in secondary electron images generated directly by ion beam imaging of the BAM L200 resolution standard and graphite pencil lead. 

\begin{figure}
	\centering
	\includegraphics[width=\textwidth]{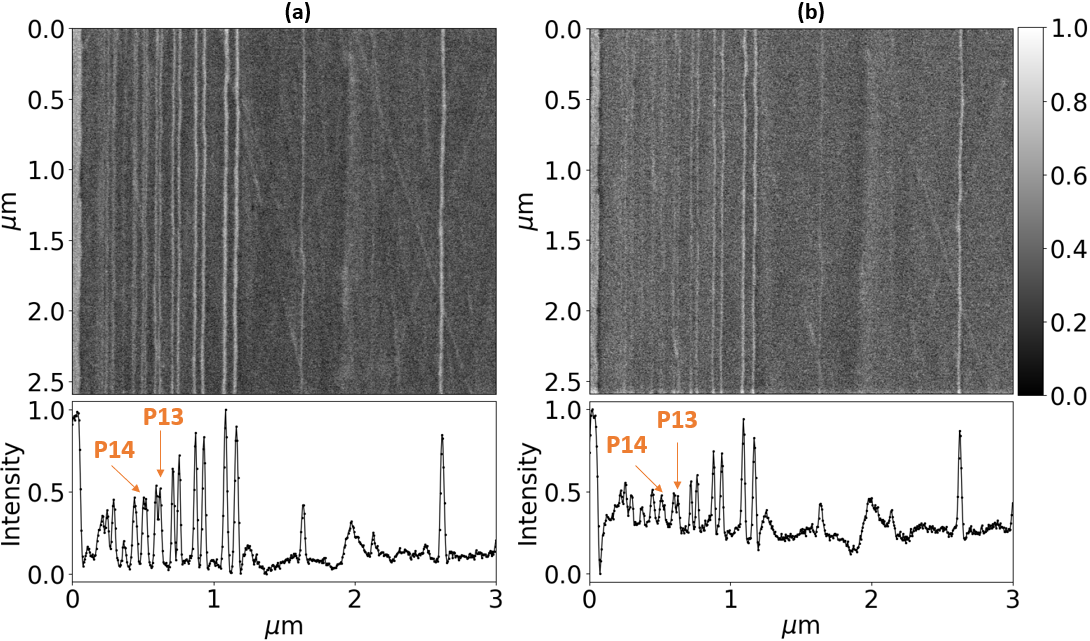}
	\caption{FIB secondary electron images of the BAM L200 resolution standard taken consecutively using a 7.5 pA, 30 keV Ga$^+$ beam and a dwell time of 500 ns per pixel. {\bf (a)} Image and line-spread function obtained by averaging the first 32 frames (corresponding to an exposure fluence of 3.5 pC/\textmu m$^2$), showing separation of period P13. The distance between the P13 strips is 11.5~nm. {\bf (b)} Image and line-spread function the same region generated by averaging a subsequent 32 frames, showing the broadening of P13 caused by ion beam intermixing of the sample.}
  \label{SIbamse}
\end{figure}

\par Figure \ref{SIbamse}a shows a FIB SE image and corresponding integrated line spread function of the BAM L200 sample obtained using a 7.5 pA, 30 keV Ga+ beam. The data were generated by averaging 32 image frames obtained using a dwell time of 500 nm per pixel. Line grating P13 is resolved and P14 is partially resolved in the image, corresponding to a d50 spot size in the range of $\sim$9-12~nm. Figure \ref{SIbamse}b shows an image an a line spread function obtained by integrating a second set of 32 image frames. It shows a resolution reduction caused by ion-induced intermixing of the sample (discussed further in the main text).

\section{Depth Resolution Calculation and Characterization}

\begin{figure}[h!]
	\centering
	\includegraphics[width=\textwidth]{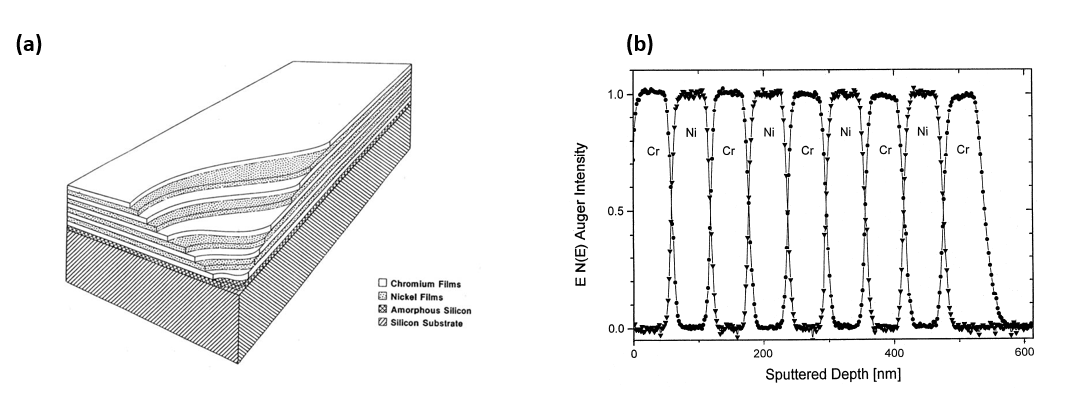}
	\caption{Structure and layout of NIST SRM 2135c certified sputter depth profiling standard of multilayered Ni/Cr alternating thin films. {\bf (a)} Sectional view of the thin-film structure. The nominal thickness of each Cr layer is 57 nm and each Ni layer is 56 nm. {\bf (b)} Normalized Auger sputter depth profile obtained using 1 keV Ar$^+$ bombardment. Images are provided with the SRM certification data sheet \cite{SRM2135c}.}
  \label{SI6}
\end{figure}

\begin{figure}[h!]
	\centering
	\includegraphics[width=\textwidth]{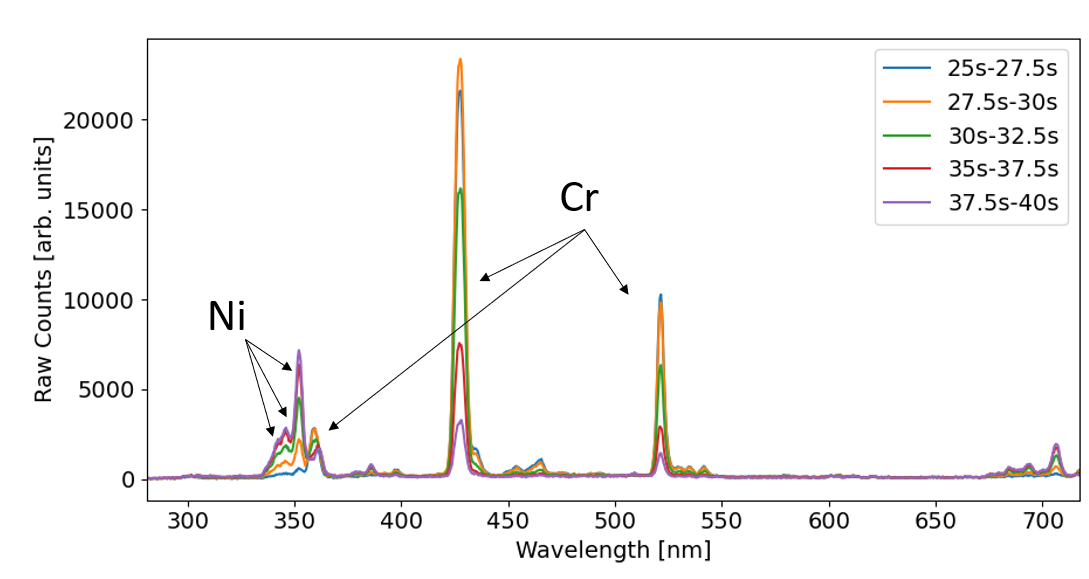}
	\caption{FIB-FS spectra of the SRM 2135c sample going from the first Cr layer into the Ni layer showing the Ni and Cr signals in raw counts from 30 keV Xe+. Each spectrum is an average of the frames over the time interval of the mill listed in the legend. The main Cr signals observed are: $3d^4(^5\text{D})4s4p(^3\text{P}^\text{o})\, \text{y}\, ^7\text{P}^\text{o}_{2,3,4}$ excited state triplet to the ground state $3d^5(^6\text{S})4s\, \text{a}\, ^7\text{S}_3$ at 357.9 nm, 359.35 nm, and 360.5 nm, the excited state triplet $3d^5(^6\text{S})4p \, \text{z}\, ^7\text{P}^\text{o}_{2,3,4}$ to the ground state at 425.4 nm, 427.5 nm, 428.97 nm, and lastly the $3d^5(^6\text{S})4p\, \text{z}\, ^5\text{P}^\text{o}_{1,2,3}$ excited state triplet to the $3d^5(^6\text{S})4s\, \text{a}\, ^5\text{S}_2$ excited state at 520.45 nm, 520.6 nm, and 520.84 nm. The main Ni signals observed are a mix of various transitions from the  $3d^9(^2\text{D})4p\, (^3\text{P}^\text{o}_{0,1,2})$ and $(^3\text{F}^\text{o}_{2,3})$ excited states to the $3d^9(^2\text{D})4s\, (^3\text{D}_{1,2,3})$ excited states between 340 nm and 352 nm. Due to the large spectral gap between the Ni and Cr signals, a large bandwidth-low resolution grating ($\sim$8 nm resolution) was used to capture both signals. The triplets and transition arrays are therefore not resolved in these spectra.}
  \label{SI7}
\end{figure}

Depth resolution measurements were performed on the NIST-certified depth profiling sample Standard Reference Material (SRM) 2135c depicted in Figure \ref{SI6}. The sample consists of alternating layers of Ni and Cr thin-films with thicknesses of 57 nm and 56 nm. Three different ion beams were tested at different landing energies using a top-down box mill and a decoupled FIB-FS spectrometer acquisition to record the Ni and Cr signals as the ion beam raster-scan milled through the layered structure. Spectra of the different layers are provided in Figure \ref{SI7} taken at different time intervals of the depth mill. Knife-edge measurements of the transition from a Ni to a Cr layer is used to determine the depth resolution. To avoid contributions from the surface oxide which can enhance the photon yield chemically \cite{Thomas1979, betz1986, betz1987, Macdonald1977} as well as to avoid the development of non-planar delayering as a result of the different sputter rates between Cr and Ni, only the first transition from Ni to Cr is considered. 

\begin{figure}[h!]
	\centering
	\includegraphics[width=\textwidth]{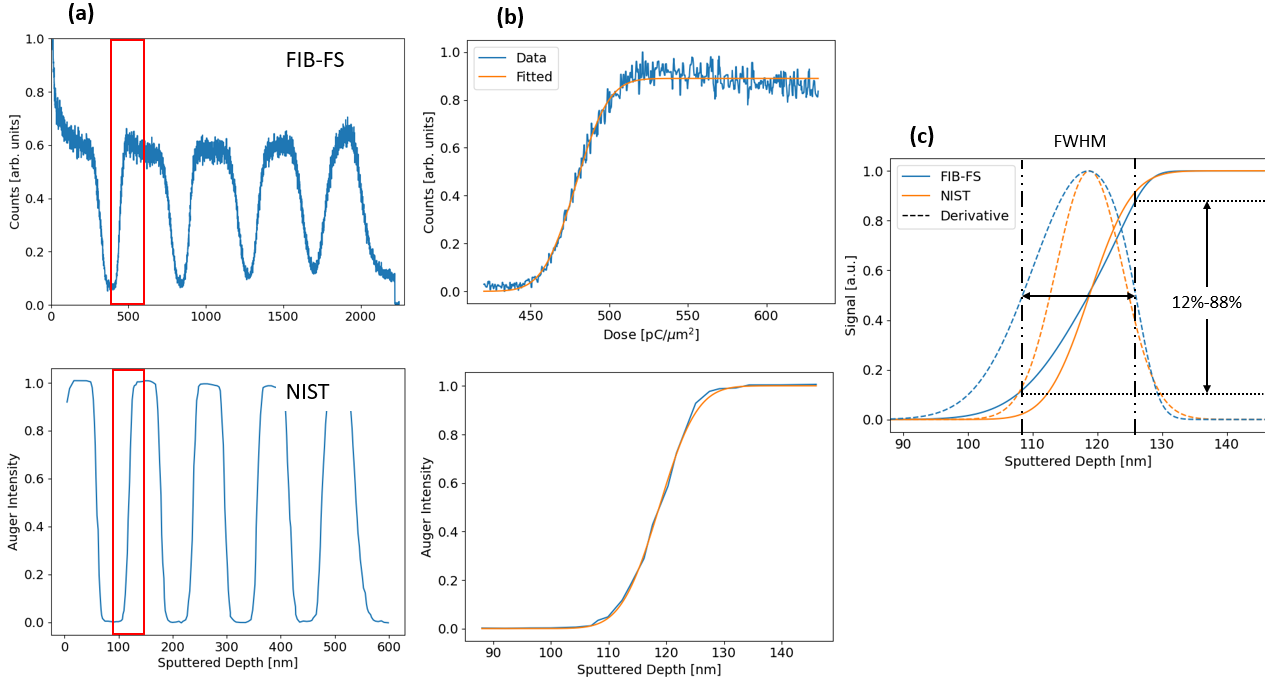}
	\caption{Method used to calculate depth resolution, demonstrated data obtained using 8 keV Ga+ ions. Resolution is determined by taking the full width at half maximum of the first derivative of the line spread function {\bf (b)} of the profile of the first Ni-to-Cr layer transition shown in {\bf (a)} and compared to the profile provided by the certificate. The full width at half maximum of the first derivative of the FIB-FS transition profile is found, corresponding to the 12\%-88\% rise time to give the resolving power of the transition in nm shown in (c).}
  \label{SI8}
\end{figure}

An overview of the analysis is shown in Figure \ref{SI8}. The Cr spectral signals are integrated, and the intensity is plotted versus time-accumulated ion dose and normalized. Since travel through the elemental layer is variable by the sputtering rate, the line spread profile must be converted from a time/ion dose domain to the spatial depth domain for comparison. The transformation is accomplished under two assumptions: that the midpoint of the Ni and Cr layer of the first transition of the FIB-FS data matches the midpoints of the Ni and Cr layers in the NIST data and that the rate of change through the layer is directly proportional to the percent change in the elemental composition by sputter rate. The sampling rate of the acquisition remains constant throughout the transition, so the transformation results in the Ni data points frequency being spread out in z-space and the Cr being compressed. For consistency of alignment to the midpoint of the layer, a fitting algorithm was used which uses a composite function of a Gaussian and square wave during the peak and valley of the transition. The second derivative of the composite fit is used to determine the edges of the transition and the midpoint between the two minimum values is used as the midpoint of the Cr and Ni layers. The data is then fitted to a standard cumulative density function (CDF) and the derivative is taken both numerically and empirically. The full width at half maximum (FWHM = 2.355$*\sigma$) is found, correlating to a 12\%-88\% rise time transition between the layers \cite{Senoner2007}. The same is repeated for the Auger sputter depth profile given in the SRM 2135c certification for comparison (see Figure \ref{SI6}).

\par
\begin{figure}
	\centering
	\includegraphics[width=0.6\textwidth]{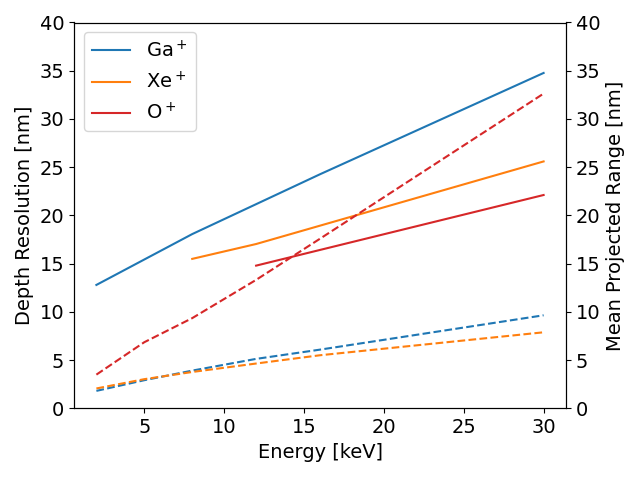}
	\caption{Measured depth resolution (solid lines) versus ion beam energy for Ga$^+$, Xe$^+$ and O$^+$ ions. Also shown is the mean projected range of each ion in Ni (dashed lines) calculated by TRIM.}
  \label{SI9}
\end{figure}

 Figure \ref{SI9} shows the results of the depth resolution measurements. Starting with 2~keV Ga$^+$, the FIB-FS depth profile resolution is 12.8~nm. This is in excellent agreement with the NIST reference measurement of 12.2~nm obtained from an Auger electron spectroscopy depth profile generated using 1~keV Ar$^+$ ions (we note that the value of $\sim$12~nm obtained at 1~keV is dominated by the true width of the interface between the Ni and Cr layers). Increasing the Ga$^+$ ion energy to 30~keV reduced the measured depth resolution to 34.8~nm, and changing the ion species to 30~keV Xe$^+$ increased it to 25.6~nm. These trends are expected since the ion range increases with energy and decreases with ion mass, unless the implanted ions are chemically active and modify the binding energies of atoms in the sample \cite{Whittmaack1990, Whittmaack1994}. Such a chemical effect is illustrated in Figure \ref{SI9} by O$^+$ ions which yield a higher depth resolution than Ga$^+$ and Xe$^+$ due to oxide formation. 

Oxide formation is illustrated further by TRIM simulations of the mean projected range of each ion in Ni (dashed lines in  Figure \ref{SI9}) --- the Ga$^+$ and Xe$^+$ data show a strong correlation between the incident ion range and the measured depth resolution. O$^+$ does not because TRIM does not account for changes in binding energy caused by oxide formation. 

\par
\begin{figure}
	\centering
	\includegraphics[width=\textwidth]{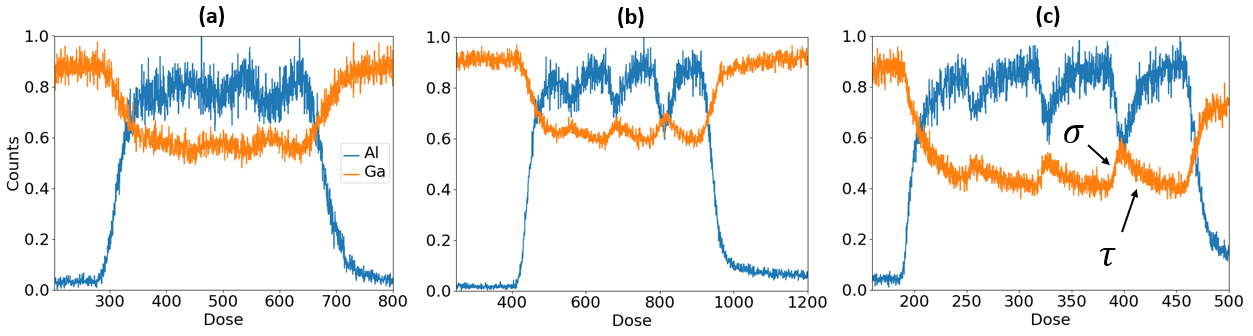}
	\caption{FIB-FS depth profiles of the quantum well structure shown in Figure 3 of the paper, taken using different ion beam energies and ion masses. {\bf (a)} Depth profiles taken with 30 keV Ga$^+$ (9 nA). The first two quantum wells (3.5 nm and 5 nm) are much thinner than the diameter of the ion interaction volume and are not resolved in the depth profiles. {\bf (b)} Depth profiles taken with 8 keV Ga$^+$ (11.2 nA). All three quantum wells are resolved. {\bf (c)} Depth profile taken with 8 keV Xe$^+$ (6 nA). All three quantum wells are resolved. The slope of the depth profile into and out of each layer, marked by $\rm \sigma$ and $\rm\tau$ respectively, are different for each ion mass and ion energy. }
  \label{mqw blowout}
\end{figure}

 The results in Figure \ref{SI9} show that depth resolution is limited by intermixing caused by ion impacts. This is illustrated further by FIB-FS depth profiles of the GaAs quantum wells in AlGaAs shown in Figure 3 of the paper, particularly if the diameter of the ion beam interaction volume is greater than the quantum well thickness. Figure \ref{mqw blowout} shows Al and Ga FIB-FS depth profiles of the sample obtained using a number of ion energies and masses. In Figure \ref{mqw blowout}a, the sample is profiled using a 30 keV Ga$^+$ beam and only the third quantum well (with a thickness of 8~nm) is resolved clearly in the depth profile. When the Ga$^+$ energy is reduced to 8~keV (Figure \ref{mqw blowout}b), and the ion mass is increased to that of Xe$^+$ (Figure \ref{mqw blowout}c), the interaction volume and the degree of intermixing decrease, and all three quantum wells are resolved in the depth profiles. The changes in ion energy and mass also alter the shapes of the depth profiles --- in particular, the slope of each 1D depth profile into and out of each quantum well layer, labelled $\rm \sigma$ and $\rm\tau$ in Figure \ref{mqw blowout}c, respectively. The changes in these slopes seen in Figure \ref{mqw blowout}a-c are caused by broadening of the interfaces through ion-solid interactions, which we discuss briefly below. 

\par There are three main attributes of ion-solid interactions that impact depth profiles obtained by ion beam techniques such as FIB-FS and SIMS: ion beam mixing, preferential sputtering, and surface texturing (topography formation). 

Topography results in increased sputtering at edges and surface features, and leads to the formation of islands/craters that expose signal from layers above/below the target plane. In most cases, this is minimised by sample preparation and cleaning methods, and by ignoring data from the periphery of the sample region irradiated by an ion beam \cite{Magee1982}. 

Preferential sputtering is the result of inherent differences in sputtering rates of different elements comprising a sample. In depth profiling, it leads to differences in endpoints for elements making up a compound, variations in perceived thicknesses of layers of differing composition measured acquired using a fixed sampling frequency, and can produce artifacts of enhanced or decreased emission at interstitial boundaries \cite{Dewitte, Kim2007, Whittmaack2001, Whittmaack1997, Zalm1995}. An example of this type of artifact is seen in the 2 keV depth profile of the quantum well structure in Figure 3 {\bf (c)} of the paper where an increase in Al signal level is seen at the leading edge of each AlGaAs layer. Here, preferential sputtering and matrix effects increase the Al signal level at each interface until a steady state is reached further in each AlGaAs layer. This effect is common in SIMS and FIB milling applications \cite{Kim2010, Havelund, Dowsett2003}, and is not unique to the FIB-FS technique.

\section{Detection Limits and Sensitivity}
Two NIST SRM samples with certified mass fractions of dozens of elements were examined using a 30 keV Xe$^+$ beam to explore the sensitivity and detection limits of the FIB-FS technique. 

\begin{figure}[h!]
	\centering
	\includegraphics[width=\textwidth]{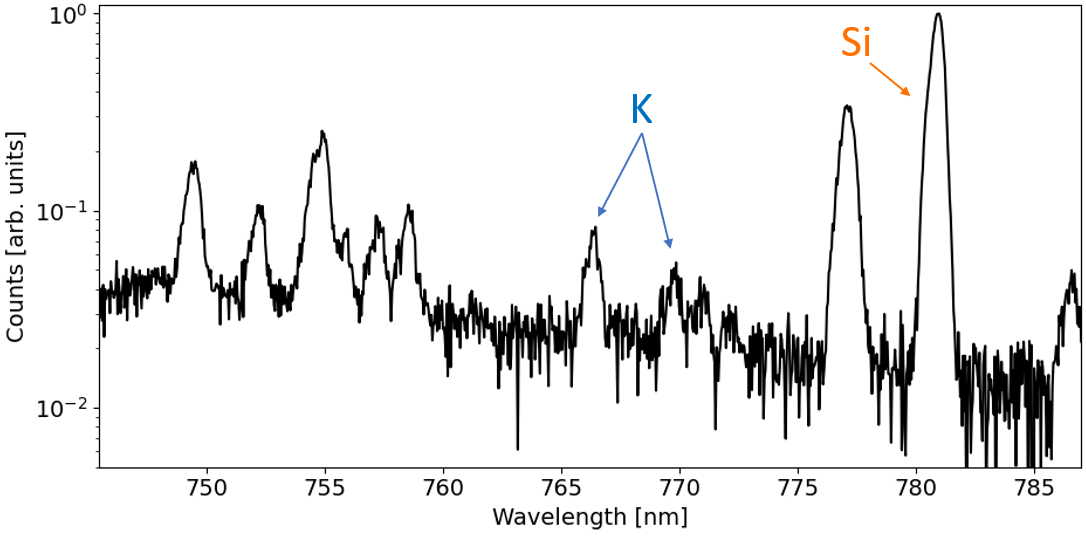}
	\caption{FIB-FS spectrum collected from NIST SRM 93a showing the detection of potassium (peaks at 766.5 nm and 769.9 nm) present at a relative concentration of 140 ppm (30 keV Xe$^+$ 375 pC/$\mu$m$^2$).}
  \label{SI12}
\end{figure}

\begin{figure}[h!]
    \centering
    \includegraphics[width=\textwidth]{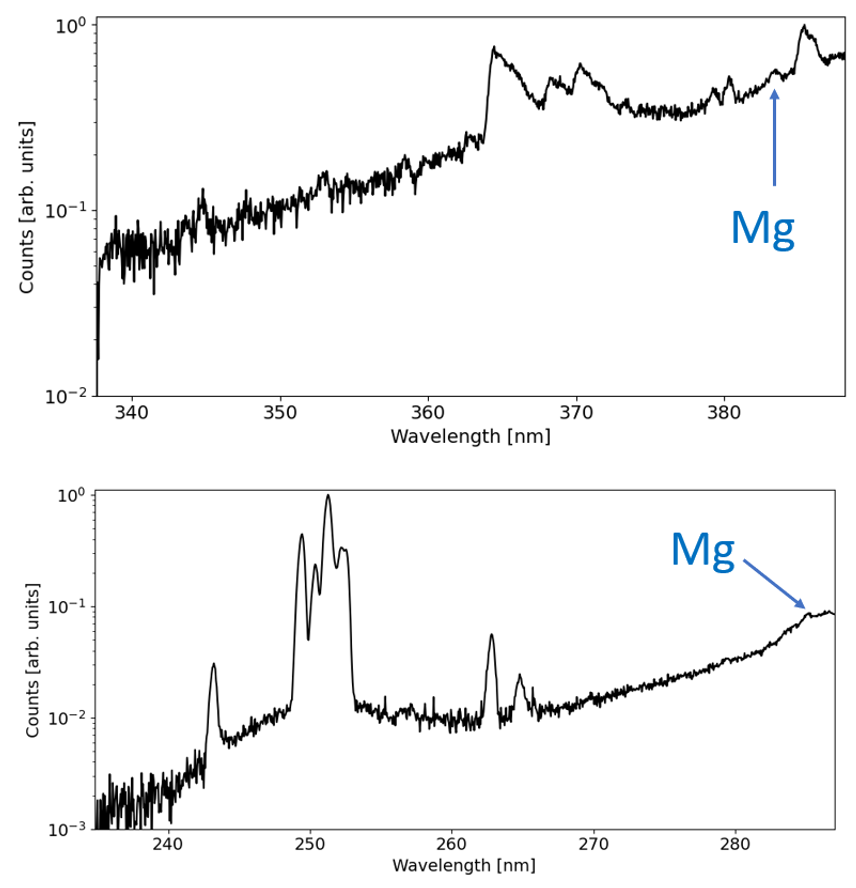}
	\caption{FIB-FS spectrum collected from NIST SRM 93a showing the detection of magnesium (peaks at 383 nm and 285 nm) present at a relative concentration of 50 ppm (30 keV Xe$^+$ 375 pC/$\mu$m$^2$).}
	\label{SI13}
\end{figure}

Reference material SRM 93a \cite{SRM93a} consists of 12 compounds embedded homogeneously in a borosilicate glass matrix. The matrix is comprised of 12.56\% B$_2$O$_3$, 80.8\% SiO$_2$, 3.98\% Na$_2$O, and 2.28\% Al$_2$O$_3$, and contains trace levels of K$_2$O and MgO which have concentrations of 0.014\% and 0.005\% by weight, respectively. FIB-FS spectra showing K and Mg emissions are shown in Figure \ref{SI12} and \ref{SI13} (bandpass filters and high-resolution gratings were employed to resolve the lines and to remove second order peaks and background emissions). Multiple peaks from each element were detected (K: $3p^64p\, (^2\text{P}^\text{o}_{\frac{3}{2},\frac{1}{2}}) \rightarrow 3p^64s\, (^2\text{S}_\frac{1}{2})$ at 766.5 nm and 769.9 nm, Mg: $3s3p\, (^1\text{P}^\text{o}_1) \rightarrow 2p^63s^2\, (^1\text{S}_0)$ at 285.2 nm and $3s3d\, (^3\text{D}_{1,2,3}) \rightarrow 3s3p\, (^3\text{P}^\text{o}_{0,1,2})$ at 382.9 nm, 383.23 nm, and 383.83 nm).

\par
\begin{figure}[h!]
    \centering
    \includegraphics[width=\textwidth]{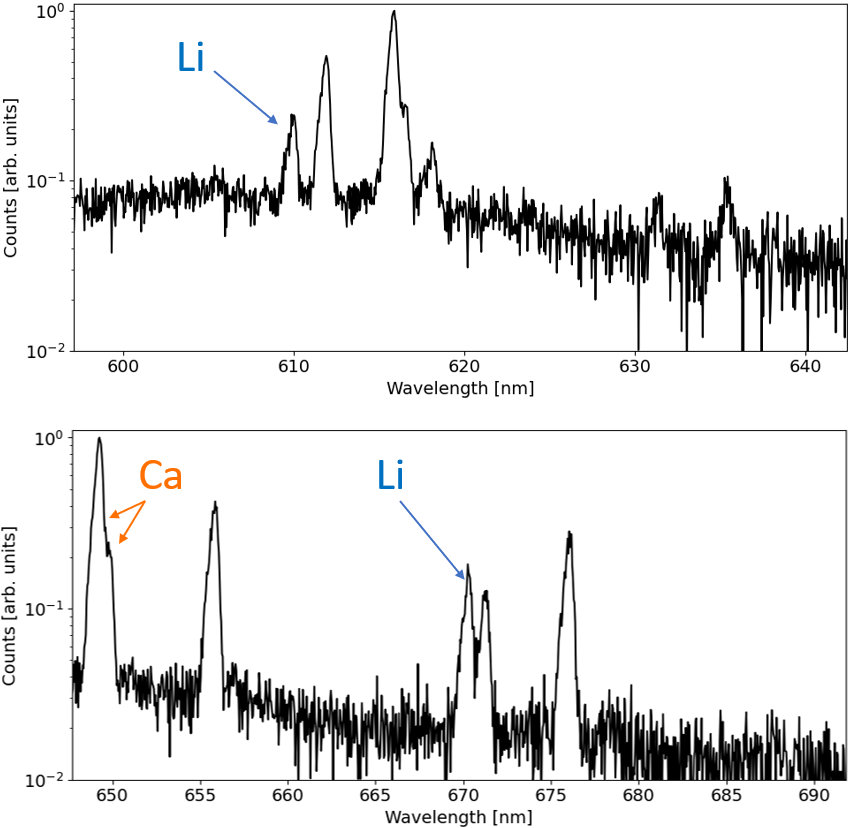}
	\caption{FIB-FS spectrum collected from NIST SRM 612 showing the detection of lithium (peaks at 610 nm and 670 nm) present at a relative concentration of 40 ppm (30 keV Xe$^+$ 37.5 and 18.7 pC/$\mu$m$^2$ for the top and bottom spectra, respectively).}
	\label{SI14}
\end{figure}

 Reference material SRM 612 \cite{SRM612} consists of 61 elements  embedded homogeneously in a glass matrix, with weight fractions in the range of 10 to 80 mg/kg. The matrix is comprised of 72\% SiO$_2$, 14\% Na$_2$O, 12\% CaO, and 2\% Al$_2$O$_3$. As in the case of SRM 93a, filters and high-resolution gratings were used to delineate emissions from specific trace elements. This sample is of interest due to the presence of Li at a concentration of $\sim 4$0~ppm. Figure \ref{SI14} shows FIB-FS spectra with lines from Li transitions  $1s^22p\, (^2\text{P}^\text{o}_{\frac{3}{2}, \frac{1}{2}}) \rightarrow 1s^22s\, (^2\text{S}_\frac{1}{2})$ at 670.77 nm and 670.8 nm and $1s^23d\, (^2\text{D}_\frac{3}{2}) \rightarrow 1s^22p\, (^2\text{P}^\text{o}_{\frac{3}{2}, \frac{1}{2}})$ at 610.35 nm and 610.36 nm.

\par To quantify these results we turn to SIMS literature, where detection limits and sensitivity are quantified by converting secondary ion counts to atomic density using calibration standards \cite{Chakraborty, Ottolini}. Following an analogous approach, the atomic concentration $n_i$ (cm$^{-3}$) of analyte $i$ in matrix $m$ can be defined by:
\begin{equation}
    n_i = \frac{P_i}{P_m} (\text{RSF})_i
\label{atomfrac}
\end{equation}
where $P_i$ and $P_m$ are the photon counts from analyte $i$ and matrix $m$, respectively. $\text{RSF}$ is a relative sensitivity factor that relates photon counts to atomic density. It can be determined from a calibration sample with a known atomic density $n_m$ (cm$^{-3}$) by comparing the ratio of photon yields $\beta$ (photons/sputtered atom) of analyte $i$ and matrix $m$ such that $\text{(RSF)} = n_m(\beta_m/\beta_i)$. Equation (\ref{atomfrac}) can then be written in terms of fractional concentrations $C_i$ and $C_m$:
\begin{equation}
    \frac{C_i}{C_m} = \frac{P_i}{P_m}\frac{\beta_m}{\beta_i}
    \label{conc}
\end{equation}
or for minimum concentrations:
\begin{equation}
    \frac{C_{i\text{(min)}}}{C_m} = \frac{P_{i\text{(min)}}}{P_m}\frac{\beta_m}{\beta_i}
    \label{climits}
\end{equation}

Hence, by using a standard of known composition, $\beta_m/\beta_i$ can be found by comparing $P_i$ to $P_m$ where $P_m$ is the photon count rate from matrix element $m$ present at fractional concentration $C_m$. 

For SRM 612, we compared the photon count rates from Li transitions  $1s^22p\, (^2\text{P}^\text{o}_{\frac{3}{2}, \frac{1}{2}}) \rightarrow 1s^22s\, (^2\text{S}_\frac{1}{2})$ at 670.77 nm and 670.8 nm to those of Ca transitions $3p^63d4p\, (^3\text{F}^\circ_2) \rightarrow 3p^63d4s\, (^3\text{D}_{1,2})$ at 649.38 and 649.96 nm. Substituting the conversion rate from equation (\ref{conc}) into (\ref{climits}), where $P_{\text{Li(min)}}$ is determined by the limit of detection defined as the sum of the limit of the background and $1.645 * \text{std}(P_{\text{Li}})$, yields $C_{\text{Li(min)}} = 0.8$ ppm. 

For SRM  93a, we compared the photon count rates from K transitions $3p^64p\, (^2\text{P}^\text{o}_{\frac{3}{2},\frac{1}{2}}) \rightarrow 3p^64s\, (^2\text{S}_\frac{1}{2})$ at 766.5 nm to that of the Si transition $3s^23p4s\, (^1\text{P}^\text{o}_1) \rightarrow 3s^23p^2\, (^1\text{S}_0)$ at 390.55 nm observed as the second order at 781.1 nm. Given the fractional concentrations $C_{\text{K}} = 0.0116\%$ and $C_{\text{Si}} = 37.77\%$, $C_{\text{K(min)}}=3.9$ ppm.

The Mg detection limit was not quantified due to the presence of the intense broadband emission seen in Figure \ref{SI13}. The spectrum is shown to illustrate qualitatively the ability of FIB-FS to detect Mg present at a concentration of 50~ppm.

\par A useful characteristic of FIB-FS is the dynamic range and modularity of the detection system compared to mass spectrometers. The spectral response can be modified with filters and gratings to preferentially increase or decrease the effective dynamic range, and a sputtered atom can emit photons at a number of wavelengths and efficiencies. This can be beneficial for comparing concentrations by providing multiple opportunities to compare matrix and analyte peaks if a particular pair is beyond the dynamic range of the detector.

\section{Li Ion Battery Analysis by EDX}

In EDX, there is a trade off between the electron beam energy needed to excite characteristic x-rays and the resulting electron-sample interaction volume which limits spatial resolution. This is one reason why FIB-FS is complimentary to EDX, as we demonstrate here using the NMC cathode characterised by FIB-FS in Figure 4 of the paper. 

\begin{figure}[h!]
	\centering
	\includegraphics[width=\textwidth]{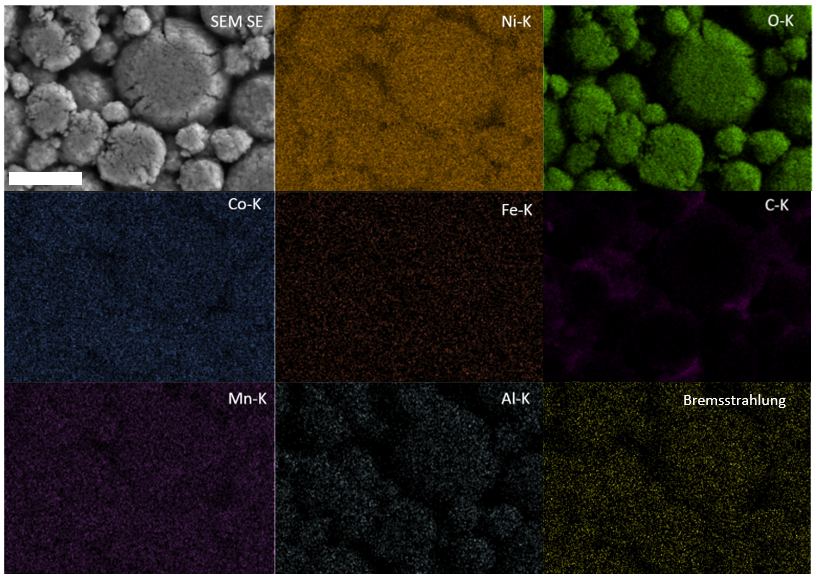}
	\caption{SEM secondary electron image and EDX elemental maps of the same region of a Li ion battery NMC cathode (573 pC/\textmu m$^2$ 20 keV electron beam). A map generated using the Bremsstrahlung x-ray background is provided for comparison. The scale bar is $\rm 20~\mu m$.}
  \label{SI15}
\end{figure}

\begin{figure}[h!]
	\centering
	\includegraphics[width=\textwidth]{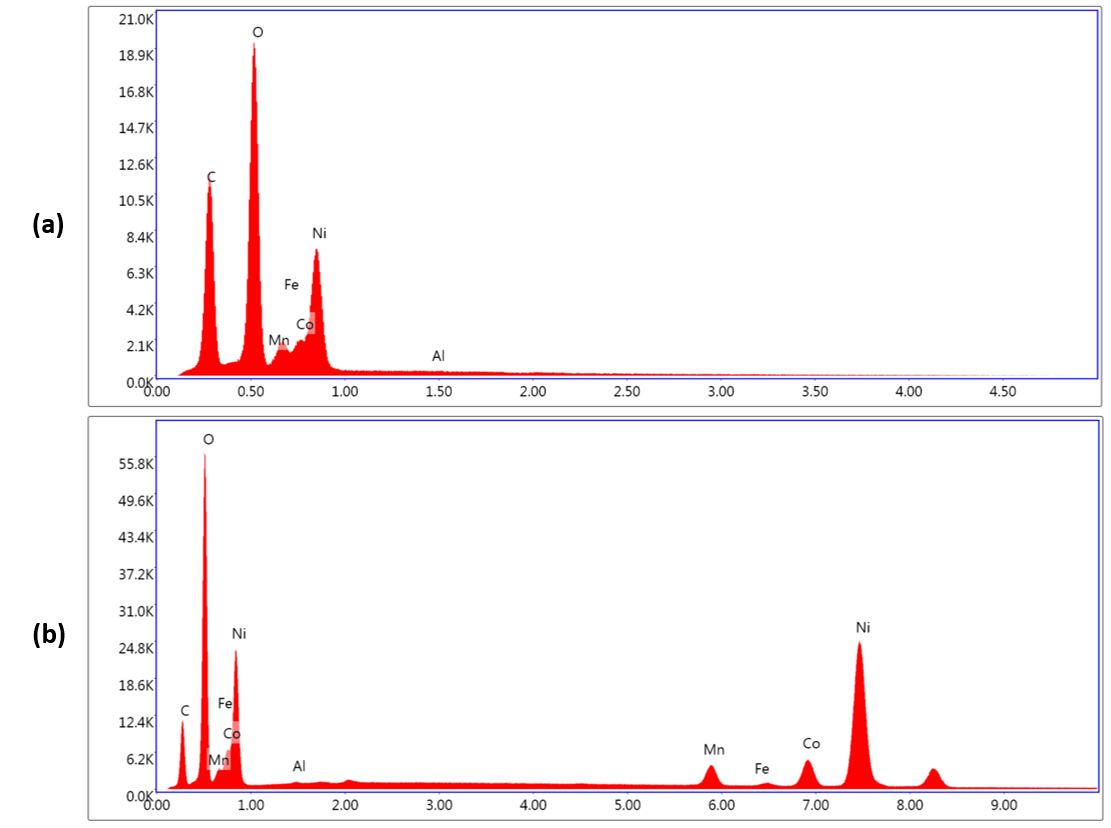}
	\caption{EDX spectra of a Li ion battery NMC cathode obtained using an electron beam energy of {\bf (a)} 5 keV and {\bf (b)} 20 keV. The electron beam energy limits the maximum energy of x-rays that can be excited in EDX.}
  \label{EDS spectra}
\end{figure}

\par Figure \ref{SI15} shows EDX elemental maps of the cathode. The maps were acquired using an EDAX Octane Elite Plus SSD detector with a silicon nitride window, using an SEM beam energy of 20 keV and a total fluence of 573 pC/$\mu$m$^2$. Compared to the FIB-FS data in Figure 4 of the paper, the spatial resolution is clearly inferior. Moreover, the trace metals Al and Fe are not detected over the Bremsstrahlung x-ray background --- in order to confirm this, we compare maps generated using x-ray counts at 1.87 keV (corresponding to Al K$\alpha$ x-rays) and 6.4 keV (Fe K$\alpha$) to a map generated using the Bremsstrahlung background at 3.0 keV. The contrast in these maps corresponds to surface topography and is not representative of the distributions of Al and Fe in the sample (which are illustrated by the FIB-FS maps in Figure 4c of the paper).

\par The above problems with EDX can sometimes be mitigated by reducing the electron beam energy. For example, reducing the energy to 5~keV results in improved EDX spatial resolution. However, a 5~keV electron beam cannot excite the K$\alpha$ x-rays above 5~keV, as is illustrated in Figure \ref{EDS spectra} by EDX spectra obtained using electron beam energies of 5~keV and 20~keV. The 5~keV spectrum does not contain Mn, Fe, Co and Ni K lines (all of which are located above 5 keV), and analysis of the corresponding L lines is complicated by spectral overlaps that are often encountered at low x-ray energies. For example, here the Mn, Fe, Co and Ni L lines are located at $\sim$0.64, 0.71, 0.78 and 0.85~KeV, whereas the resolution of EDX detectors is typically on the order of 100~eV. 

Peak overlaps are also not uncommon at higher energies. For example, in Figure \ref{EDS spectra}b, the Fe K$\alpha$ line at 6.4~keV overlaps with and cannot be resolved from the Mn K$\beta1$ transition at 6.5~keV. 

We note that windowless EDX detectors improve low energy performance and enable the detection of Li. However, peak overlaps are nonetheless common in multi-element materials and EDX is fundamentally unable to detect hydrogen, a ubiquitous element that affects optical, electronic and physical properties of many materials.

\section{FIB-FS Analysis Modes and Post-Processing}

\begin{figure}[h!]
	\centering
	\includegraphics[width=\textwidth]{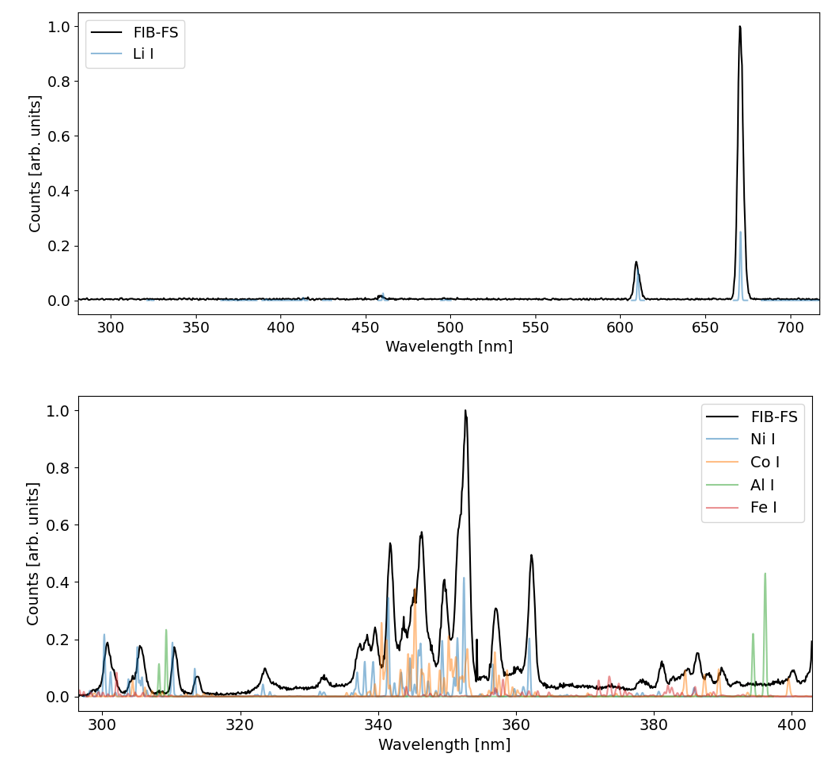}
	\caption{Spectra of the NMC cathode collected in Survey Mode.}
  \label{SI19}
\end{figure}

Here we use the NMC cathode sample to outline the FIB-FS analysis workflow used in this work.

\par
\begin{figure}[h!]
    \centering
    \includegraphics[width=\textwidth]{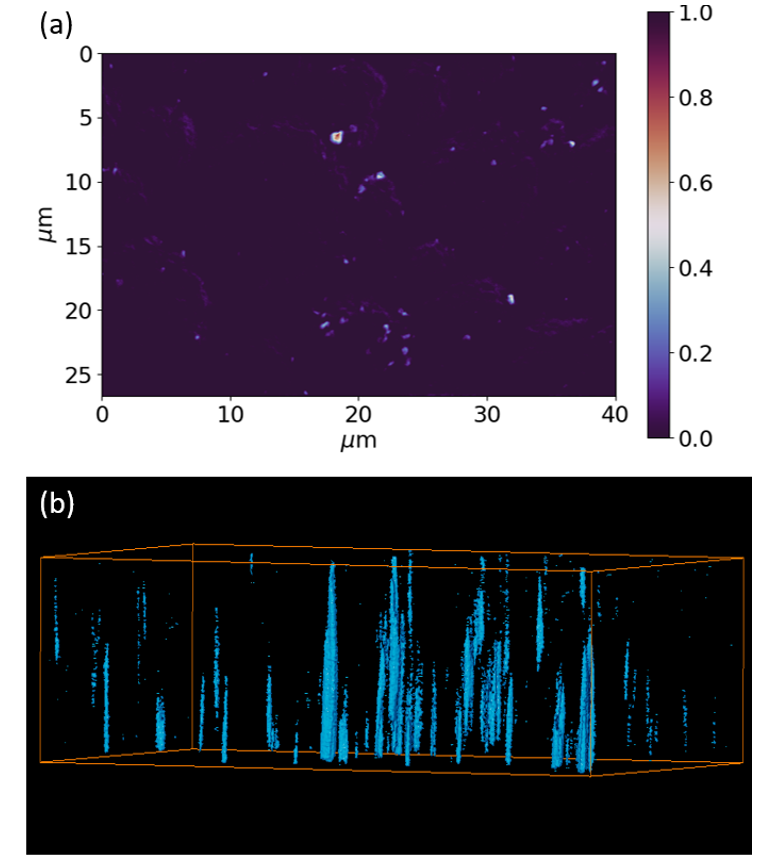}
	\caption{FIB-FS image illustrating Mg dopant distributions in an NMC lithium-ion battery cathode (30~keV Ga$^+$). {\bf (a)} 2D map of the Mg dopant distribution (1.3 nC/\textmu m$^2$). {\bf (b)} 3D volume reconstruction of the Mg distribution in the sample (2.7 pC/\textmu m$^2$ per frame).}
	\label{SI20}
\end{figure}

 A sample is initially analyzed in Survey Mode, where we raster scan the ion beam using the standard microscope control software. An elliptical mirror and an EMCCD camera are used to collect the FIB-FS signal. Various gratings are used based on which elements are expected in the sample or observed in the spectra. This is illustrated by Figure \ref{SI19} by two spectra acquired using gratings blazed at 500 and 300~nm in order to efficiently detect a Li emission at $\sim 671$~nm, and Ni, Co and Mn emissions located between 300 and 400~nm. However, in these spectra, there is no clear indication of Al or Fe signals expected from this sample. This is a consequence of the fact that Fe and Al are present in trace quantities, their distributions are highly heterogeneous, and the ion beam scan is decoupled from FIB-FS signal acquisition. To overcome this problem, the signal collection period is coupled to the ion beam scan such that a spectrum is collected at each pixel, yielding the second two modes of operation: Hyperspectral Mode and Photon Imaging Mode. When this process is repeated multiple times, ion beam sputtering yields an image stack that can be used to generate 2D maps and a 3D volumetric depth profile. This process is demonstrated by Figure \ref{SI20} for the Mg dopant in the NMC sample.

\begin{figure}[h!]
    \centering
    \includegraphics[width=\textwidth]{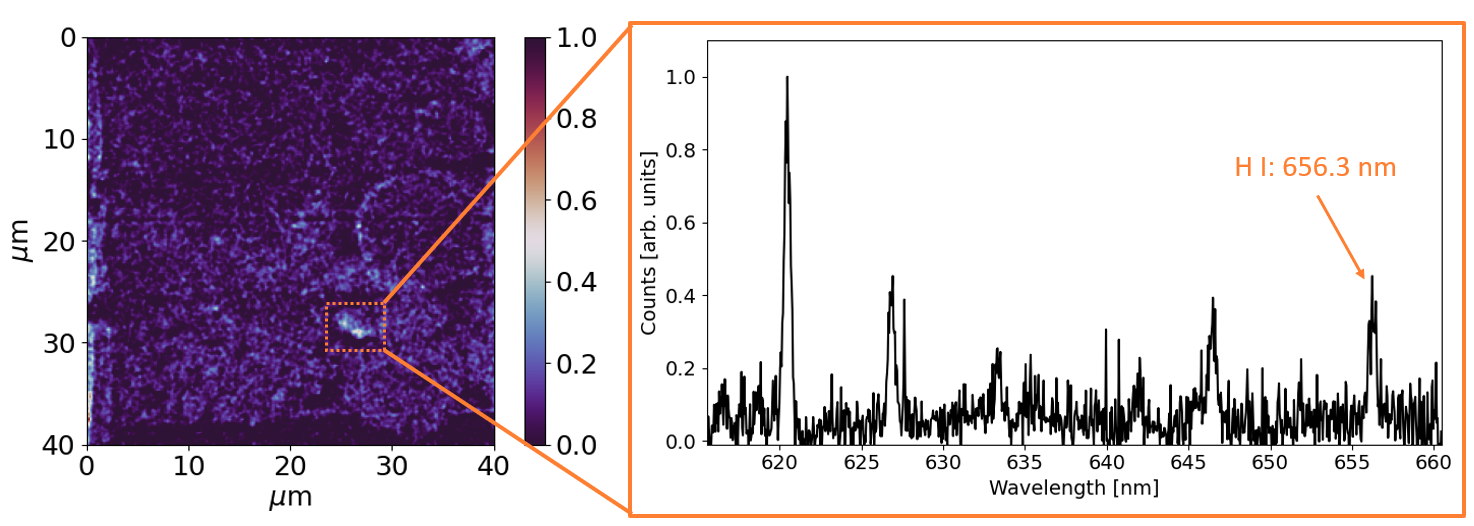}
	\caption{FIB-FS Hyperspectral map of the NMC cathode and a spectrum showing clearly a hydrogen emission at 656.3~nm.}
	\label{SI21}
\end{figure}

\par Additional analysis of specific pixels is done to confirm conclusions about local variations in composition. For example, spectra from pixels corresponding to a region of interest can be averaged, as is illustrated by Figure \ref{SI21} where the spectra contained in the orange box are averaged to reveal a hydrogen peak located at 656.3~nm.

\section{Ga Ion Beams and FIB-FS Spectral Overlaps}
\begin{figure}[h!]
    \centering
    \includegraphics[width=\textwidth]{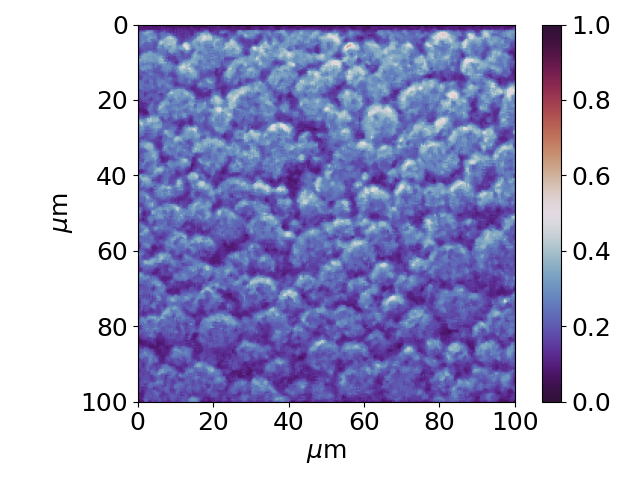}
	\caption{FIB-FS image of an NMC cathode taken using an O$^+$ plasma FIB showing the Mn distribution. The Mn distribution is similar to that of Ni and Co shown in Figure 4c of the paper (as expected).}
	\label{SI22}
\end{figure}

Most commercial FIB instruments employ liquid metal Ga+ sources. Consequently, the Ga+ ions are implanted in, and re-sputtered and backscattered from samples. In FIB-FS, sputtered Ga emits photons due to transitions from the $4s^25s\, (^2\text{S}_\frac{1}{2})$ excited state to the $3d^{10}4s^24p\, (^2\text{P}^\text{o}_{\frac{1}{2}, \frac{3}{2}})$ ground state, located at 403.3~nm and 417.2~nm. In most cases, these emissions can be ignored. However, in the case of Mn, the most efficient emissions are at 403.1~nm, 403.31~nm and 403.5~nm (due to the transition from the triplet state $3d^5(^6\text{S})4s4p(^3\text{P}^\text{o})\, \text{z}\, ^6\text{P}^\text{o}_{\frac{3}{2}, \frac{5}{2}, \frac{7}{2}}$  to the ground state $3d^54s^2\, (\text{a}\, ^6\text{S}_\frac{5}{2})$). Since our setup did not have the spectral resolution needed to resolve these peaks, we instead use a different primary ion source to eliminate the Ga+ emission. Figure \ref{SI22} shows FIB-FS map generated using the Mn signal at 403 nm, excited using an O+ plasma FIB.

\clearpage

\bibliography{SI_bibfile1}